%% file: main.tex
\documentclass[fleqn,usenatbib]{mnras}

\usepackage{newtxtext,newtxmath,comment}

\DeclareRobustCommand{\VAN}[3]{#2}
\let\VANthebibliography\thebibliography
\def\thebibliography{\DeclareRobustCommand{\VAN}[3]{##3}\VANthebibliography}

\usepackage[utf8]{inputenc}
\usepackage[T1]{fontenc}
\usepackage{enumitem}
\usepackage{lipsum,hyperref,afterpage,caption,subcaption,float}
\hypersetup{ 
			colorlinks=true,
			pdfauthor={Lamman, Claire},
			pdftitle={desiIA-draft},
			citecolor=blue
			}

\usepackage{graphicx}	
\usepackage{amsmath}	
\usepackage{lineno}
\let\oldequation\equation
\let\oldendequation\endequation
\renewenvironment{equation}
  {\linenomathNonumbers\oldequation}
  {\oldendequation\endlinenomath}
\usepackage[dvipsnames]{xcolor}
\usepackage{orcidlink}

\title[Detection of galaxy multiplet alignment]{Detection of the large-scale tidal field with galaxy multiplet alignment in the DESI Y1 spectroscopic survey}

\author[Claire Lamman]{Claire Lamman \orcidlink{0000-0002-6731-9329},$^{1}$\thanks{E-mail: claire.lamman@cfa.harvard.edu}
Daniel Eisenstein,$^{1}$  
Jaime E. Forero-Romero \orcidlink{0000-0002-2890-3725},$^{2,3}$
Jessica Nicole Aguilar,$^{4}$\newauthor
Steven Ahlen \orcidlink{0000-0001-6098-7247},$^{5}$
Stephen Bailey \orcidlink{0000-0003-4162-6619},$^{4}$
Davide Bianchi \orcidlink{0000-0001-9712-0006},$^{6}$
David Brooks,$^{7}$
Todd Claybaugh,$^{4}$\newauthor
Axel de la Macorra \orcidlink{0000-0002-1769-1640},$^{8}$
Peter Doel,$^{7}$
Simone Ferraro \orcidlink{0000-0003-4992-7854},$^{4,9}$
Andreu Font-Ribera \orcidlink{0000-0002-3033-7312},$^{7,10}$\newauthor
Enrique Gaztañaga,$^{11,12,13}$
Satya Gontcho A Gontcho \orcidlink{0000-0003-3142-233X},$^{4}$
Gaston Gutierrez,$^{14}$
Klaus Honscheid,$^{15,16,17}$\newauthor
Cullan Howlett \orcidlink{0000-0002-1081-9410},$^{18}$
Anthony Kremin \orcidlink{0000-0001-6356-7424},$^{4}$
Andrew Lambert,$^{4}$
Martin Landriau \orcidlink{0000-0003-1838-8528},$^{4}$
Laurent Le Guillou \orcidlink{0000-0001-7178-8868},$^{19}$\newauthor
Michael E. Levi \orcidlink{0000-0003-1887-1018},$^{4}$
Aaron Meisner \orcidlink{0000-0002-1125-7384},$^{20}$
Ramon Miquel,$^{21,10}$
John Moustakas \orcidlink{0000-0002-2733-4559},$^{22}$
Jeffrey A. Newman \orcidlink{0000-0001-8684-2222},$^{23}$\newauthor
Gustavo Niz \orcidlink{0000-0002-1544-8946},$^{24,25}$
Francisco Prada \orcidlink{0000-0001-7145-8674},$^{26}$
Ignasi P\'erez-R\`afols \orcidlink{0000-0001-6979-0125},$^{27}$
Ashley J. Ross \orcidlink{0000-0002-7522-9083},$^{15,28,17}$\newauthor
Graziano Rossi,$^{29}$
Eusebio Sanchez \orcidlink{0000-0002-9646-8198},$^{30}$
Michael Schubnell,$^{31,32}$
David Sprayberry,$^{20}$
Gregory Tarl\'e \orcidlink{0000-0003-1704-0781},$^{32}$\newauthor
Mariana Vargas-Magaña \orcidlink{0000-0003-3841-1836},$^{8}$
Benjamin Alan Weaver,$^{20}$
Hu Zou \orcidlink{0000-0002-6684-3997}$^{33}$
\\ \\
Affiliations are listed at the end of the paper}

\date{Accepted XXX. Received YYY; in original form ZZZ}

\pubyear{2024}

\begin{document}
\label{firstpage}
\pagerange{\pageref{firstpage}--\pageref{lastpage}}
\maketitle

\begin{abstract}
We explore correlations between the orientations of small galaxy groups, or ``multiplets'', and the large-scale gravitational tidal field. Using data from the Dark Energy Spectroscopic Instrument (DESI) Y1 survey, we detect the intrinsic alignment (IA) of multiplets to the galaxy-traced matter field out to separations of $100 h^{-1}$Mpc. Unlike traditional IA measurements of individual galaxies, this estimator is not limited by imaging of galaxy shapes and allows for direct IA detection beyond redshift $z=1$. Multiplet alignment is a form of higher-order clustering, for which the scale-dependence traces the underlying tidal field and amplitude is a result of small-scale ($<1h^{-1}$Mpc) dynamics. Within samples of bright galaxies (BGS), luminous red galaxies (LRG) and emission-line galaxies (ELG), we find similar scale-dependence regardless of intrinsic luminosity or colour. This is promising for measuring tidal alignment in galaxy samples that typically display no intrinsic alignment. DESI's LRG mock galaxy catalogues created from the A\textsc{bacus}S\textsc{ummit} N-body simulations produce a similar alignment signal, though with a 33\% lower amplitude at all scales. An analytic model using a non-linear power spectrum (NLA) only matches the signal down to 20$h^{-1}$Mpc. Our detection demonstrates that galaxy clustering in the non-linear regime of structure formation preserves an interpretable memory of the large-scale tidal field. Multiplet alignment complements traditional two-point measurements by retaining directional information imprinted by tidal forces, and contains additional line-of-sight information compared to weak lensing. This is a more effective estimator than the alignment of individual galaxies in dense, blue, or faint galaxy samples.\\

\end{abstract}




\section{Introduction}

Galaxies form and reside within a large-scale structure primarily composed of dark matter. This spatial clustering is shaped by gravitational forces acting on initially small perturbations present in the very early universe. As structure grows hierarchically through gravitational instability, the tidal fields associated with the evolving matter density are expected to induce subtle effects on the shapes, spins, and orientations of galaxies and dark matter haloes.

These correlations are broadly known as "Intrinsic Alignments" (IA). Generally, elliptical galaxies and haloes display a linear relationship with the large-scale tidal field, where long axes are aligned with its stretching direction. For a pedagogical introduction to IA, see \cite{lamman_ia_2023} and for comprehensive reviews, see \cite{joachimi_galaxy_2015} and \cite{troxel_intrinsic_2015}. IA are most commonly studied as a contaminant of cosmological probes, such as weak lensing and redshift-space distortions (RSD), but in principle they can also be used to trace any cosmological effect which is imprinted in the large-scale density field \citep{chisari_cosmological_2013}. Compared to traditional two-point clustering statistics, IA have the advantage of capturing both the magnitude and polarization of tidal shear, as is done with weak lensing. While weak lensing traces all foreground matter, IA from spectroscopic data contain additional information along the line-of-sight. However, the effect is subtle and requires large samples and high-quality imaging. IA have been explored as a probe of primordial non-gaussianity \citep{akitsu_imprint_2021, kurita_constraints_2023}, Baryon Acoustic Oscillations \citep{okumura_intrinsic_2019, xu_evidence_2023}, Redshift-space Distortions \citep{okumura_first_2023}, and cosmic B-modes \citep{georgiou_b-modes_2023, akitsu_gravitational_2023, saga_imprints_2024}. 

In some cases it is advantageous to study the alignment of galaxy ensembles: groups and clusters as opposed to individuals. The determined shapes of galaxy ensembles are unaffected by the myriad of systematic effects which arise from imaging, and are associated with the shape of their host haloes, which display stronger tidal alignment \citep{smargon_detection_2012, fortuna_halo_2021, lee_disentangling_2023}. Clusters of Luminous Red Galaxies (LRGs) in the Sloan Digital Sky Survey display similar but stronger alignment compared to single galaxies \citep{smargon_detection_2012, van_uitert_intrinsic_2017}. These correlations were found to be lower than predicted by N-body simulations, which may be due to hydrodynamic or projection effects, which create misidentification of cluster members \citep{shi_intrinsic_2024}. There are also concerns of orientation bias in identifying clusters, particularly for photometric surveys \citep{sunayama_observational_2023}.

\begin{figure} 
\centering
\includegraphics[width=.48\textwidth]{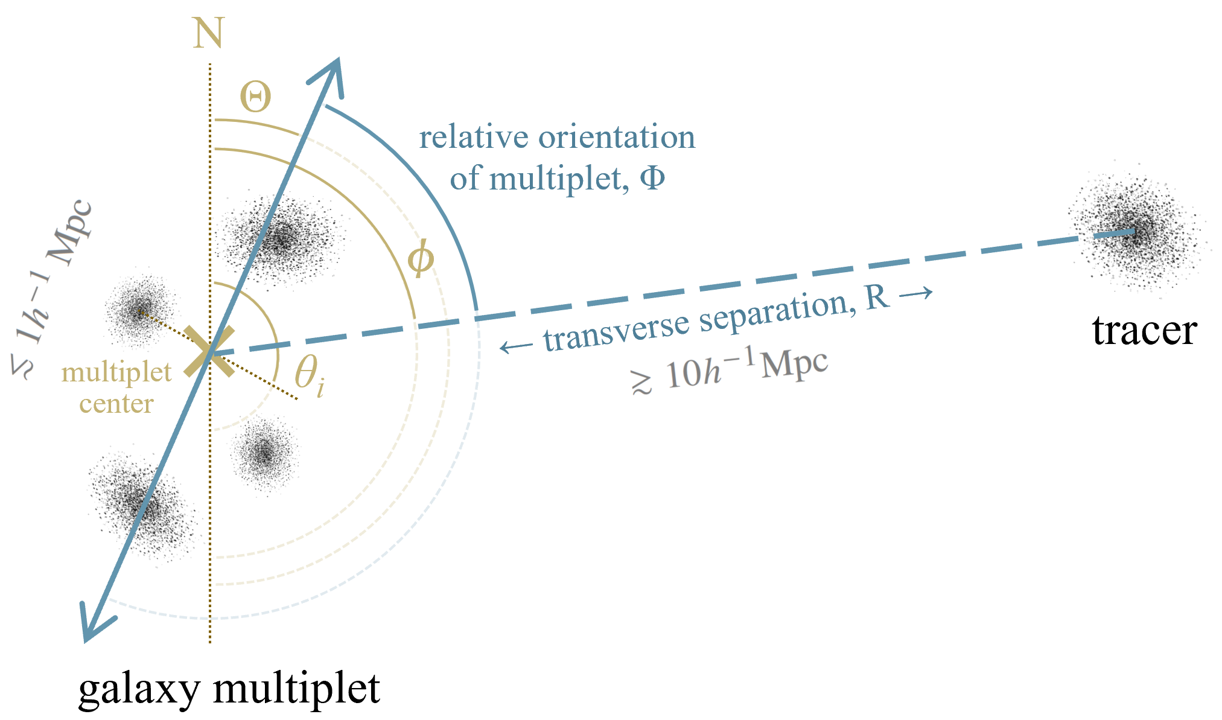}
\caption{A schematic showing the two parameters of our multiplet alignment estimator: the projected orientation $\Phi$ of the multiplet relative to a tracer, and their projected separation, $R$. The variables used to determine these, as described in Section~\ref{sec:formalism}, are displayed in gold and shown relative to North. They are: the position angle of multiplet members, shown with one $\theta_i$, the reduced multiplet orientation $\Theta$, and the position angle of the tracer relative to the multiplet $\phi$.}
\label{fig:schematic}
\end{figure}

In this work we explore the potential of using galaxy ``multiplets'': small sets of galaxies, mostly consisting of 2-4 members within 1 $h^{-1}$Mpc of each other (Fig.~\ref{fig:schematic}). We expect these tiny ensembles to still preserve information from the large-scale tidal field, while being more abundant than larger groups. Multiplets are not necessarily virialized systems, but can be understood in the IA framework as they are well within the nonlinear regime of gravitational evolution. Like galaxy shapes and haloes, their orbital structure carries a memory of the initial tidal field. 

The alignment of galaxy multiplets may be a better estimator than individual galaxies when: imaging is poor, the sample is especially dense, or the sample displays little or no individual alignment, as is the case for spiral (or ``blue'') galaxies. The latter of these applies to most available spectroscopic samples beyond redshift 1. Understanding the redshift evolution of IA is an important component of fully utilising forthcoming cosmic shear surveys \citep{dark_energy_survey_and_kilo-degree_survey_collaboration_y3_2023}. However the redshift evolution of IA is unclear and there is no direct IA detection beyond redshift 1 with traditional estimators. 

We describe and model this estimator from the perspective of IA, but this work is also related to the fields of both galaxy groups and higher-order clustering. Although multiplets are not galaxy groups, which are virialized systems and typically describe more complete sets of galaxies \citep{oppenheimer_simulating_2021}, multiplets exist on similar scales. They can overlap group catalogues, especially when multiplets are identified in dense samples. Furthermore, the nonlinear dynamics within groups directly affect the amplitude of multiplet alignment. 

Since, in most cases, we are measuring the orientation of close galaxy pairs relative to a distant tracer, this estimator can also be thought of as a squeezed three-point correlation function. Previous work has explored 3-point and higher-order correlations in spectroscopic data \citep{slepian_computing_2015, philcox_encore_2022}, including detecting evidence of the tidal field \citep{slepian_detection_2017} and investigating the squeezed 3-point function \citep{yuan_using_2017}. These describe correlations that arise from larger scales than multiplets, but are a similar framing of our estimator.

As a spectroscopic survey of over 40 million galaxies, the DESI Survey (Dark Energy Spectroscopic Instrument), is well-suited to probing subtle, higher-order clustering effects in three dimensions \citep{levi_desi_2013, desi_collaboration_desi_2016, desi_collaboration_desi_2016-1, desi_collaboration_overview_2022, desi_collaboration_early_2023, miller_optical_2023}. To explore the potential of multiplet IA, we measure the tidal alignment of multiplets in DESI's Year 1 survey \citep{DESI2024.III.KP4, DESI2024.IV.KP6, DESI2024.VI.KP7A}. We use three galaxy samples: bright galaxies (BGS), luminous red galaxies (LRG), and emission-line galaxies (ELG), ranging from redshifts 0.1 -- 1.5. As a proof-of-concept for interpreting this estimator, we develop modelling for the catalogue which displays the highest galaxy bias and alignment signal, LRGs.

Section~\ref{sec:desi_catalogue} describes the DESI data and mock catalogues used. Section~\ref{sec:method} outlines our methodology for identifying galaxy multiplets and measuring their alignment. Section~\ref{sec:interp} presents a comparison to mock catalogues and an analytic model of the alignment signal. Section~\ref{sec:conclusion} summarizes key results and discusses prospects for utilising future datasets.

Throughout the paper we assume the cosmological parameters of $H_0=69.6$, $\Omega_{m,0}=0.286$, $\Omega_{\Lambda,0}=0.714$.

\begin{figure*} 
\includegraphics[width=\textwidth]{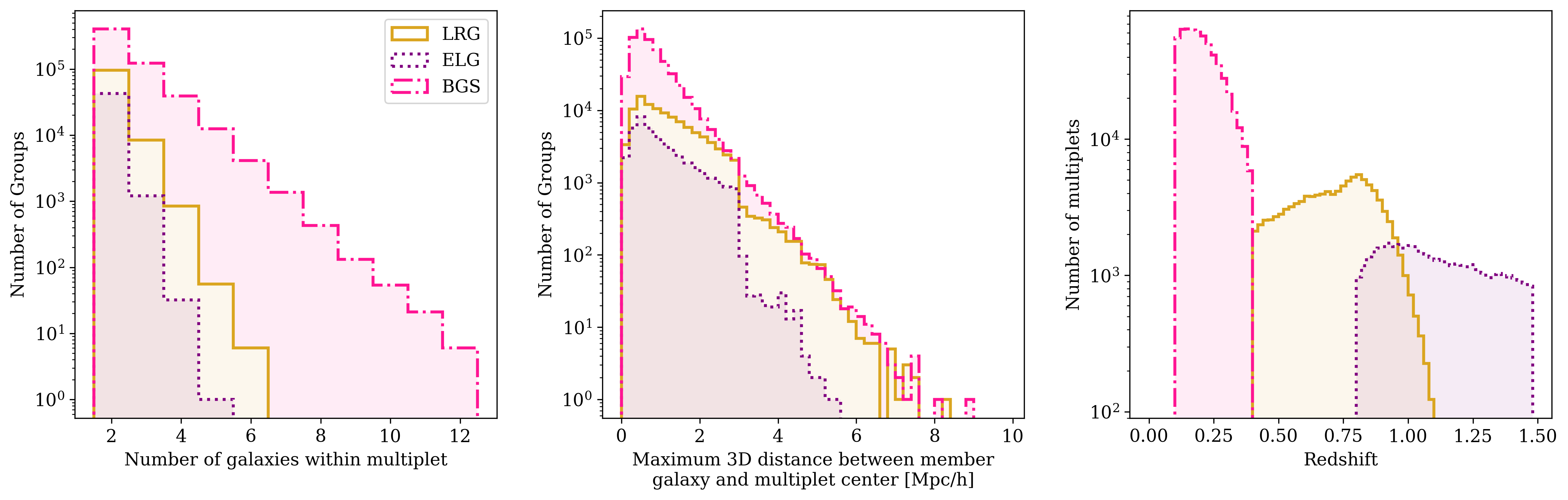}
\caption{Demographics of multiplets found in the galaxy catalogues of DESI's Year 1 survey. The majority of multiplets in all cases are composed of only two members, even for the densest sample, BGS, where 70\% of multiplets are galaxy pairs. The spatial size of multiplets is shown in the middle panel, which is described by the maximum 3D distance between a multiplet member and the multiplet's centre in redshift space. There is a drop around 3 $h^{-1}$Mpc, corresponding to the maximum distance to the center for a pair of galaxies based on our multiplet definition (Section~\ref{sec:id_multiplets}). The right panel shows the redshift distribution of multiplets.}\label{fig:demographics}
\end{figure*}

\section{DESI catalogues}\label{sec:desi_catalogue}

DESI's targets are chosen from DR9 of the Legacy Imaging Survey \citep{dey_overview_2019, myers_target-selection_2023}. For more information on DESI's target selection, see \cite{desi_collaboration_early_2023, desi_collaboration_validation_2023}. We used spectroscopic redshifts from DESI's Year 1 data \citep{guy_spectroscopic_2023, schlafly_survey_2023}. This data will be publicly available with DESI's Data Release 1 (DR1) \citep{DESI2024.I.DR1}, and documented in \citet{DESI2024.II.KP3}. The catalogues we use are designed for measuring large-scale structure \citep{KP3s15-Ross}. They contain spectra of 3.3 million BGS within $0.1<z<0.4$, 2.2 million LRGs within $0.4<z<1.1$, and 2.7 million ELG within $0.8<z<1.5$. Note that this is DESI's full BGS catalogue, as opposed to the luminosity-limited sample used for BAO analysis \citep{DESI2024.III.KP4}. More information on these selection and validation of these samples can be found in \cite{hahn_desi_2023, zhou_target_2023, raichoor_target_2023}. The catalogues also include weights to account for redshift failure and the probability that each target was observed. When making our measurement, we applied these weights to all tracer samples.

For exploring the difference between red and blue galaxy populations, we create two sub-catalogues of the BGS sample within $0.1<z<0.2$. The BGS Blue sample is defined as galaxies with a $g-r$ colour of less than $0.3 + 3z$, and BGS Red as galaxies with a $g-r$ colour greater than $0.5 + 3z$. The colour $g-r$ is computed from the Legacy Survey imaging and $z$ is the galaxies' spectroscopic redshifts.

Validation of DESI's survey selection and analysis rely on mock catalogues. These are generated with 25 N-body cosmological simulations from A\textsc{bacus}S\textsc{ummit} \citep{hadzhiyska_compaso_2021, maksimova_abacussummit_2021, yuan_redshift_2024}. An important aspect of these is to capture the effects of fiber assignment. DESI performs spectroscopy on thousands of objects at once via individually-controlled robotic positioners, which place fiber optic cables on galaxies. Their positions are limited to a set patrol radius and by proximity to other fibers, resulting in an under-sampling of highly clustered targets. Here we use the set of mock catalogues prepared for DESI Y1 clustering measurements, and two implementations of fiber assignment. The first is aMTL (alternative Merged Target Ledgers), where the probability that targets are observed is estimated by running the fiber assignment algorithm with varying target priorities \citep{lasker_production_2024}. The second is FFA (Fast Fiber Assignment), which determines the probability that galaxies are observed based on the number of survey passes at its location and number nearby galaxies (Bianchi et al., in prep).

\begin{table}
\begin{centering}
\begin{tabular}{|c|c|r|r|r|}
\hline
\multicolumn{1}{|c|}{\textbf{Galaxy}} & \multicolumn{1}{c|}{\textbf{Redshift}} & \multicolumn{1}{c|}{\textbf{N}} & \multicolumn{1}{c|}{\textbf{N galaxy}} & \multicolumn{1}{c|}{\textbf{Volume}} \\
\multicolumn{1}{|c|}{\textbf{type}} & \multicolumn{1}{c|}{\textbf{Range}} & \multicolumn{1}{c|}{\textbf{galaxies}} & \multicolumn{1}{c|}{\textbf{multiplets}} & \multicolumn{1}{c|}{\textbf{[Gpc$^3h^{-3}$]}}\\ \hline
ELG & 1.1 < z < 1.5 & 1.5 M & 21  K & 67.8 \\ 
ELG & 0.8 < z < 1.1 & 1.2 M & 22  K & 35.8 \\ \hline
LRG & 0.8 < z < 1.1 & 0.9 M & 34 K & 34.6  \\
LRG & 0.4 < z < 1.1 & 2.2 M & 105 K & 34.6  \\ \hline
BGS & 0.3 < z < 0.4 & 0.6 M & 64  K & 3.2 \\
BGS & 0.2 < z < 0.3 & 1.3 M & 212 K & 1.5 \\
BGS & 0.1 < z < 0.2 & 1.4 M & 307 K & 0.5 \\
BGS Blue & 0.1 < z < 0.2 & 0.56 M & 81 K & 0.5 \\
BGS Red & 0.1 < z < 0.2 & 0.54 M & 100 K & 0.5 \\
\hline
\end{tabular}\caption{Properties of the DESI catalogues used to identify galaxy multiplets. The right column shows the comoving volume of the sample, estimated from the positions of galaxy multiplets. The colour cuts used to make the BGS Blue and Red samples are described in Section~
\ref{sec:desi_catalogue}.}\label{tab:samples}
\end{centering}
\end{table}

\section{Alignment method}\label{sec:method}

\subsection{Identifying galaxy multiplets}\label{sec:id_multiplets}
Our measurement is a projected quantity, relating the orientation of multiplets in the plane of the sky as a function of transverse distance (Figure \ref{fig:schematic}). However, we identify small multiplets of galaxies in 3D comoving space using spectroscopic redshifts. Each galaxy is matched to its nearest neighbour and all pairs are limited to a maximum separation in the plane of the sky, $r_p$, and along the line of sight, $r_\parallel$. $r_\parallel$ is necessarily larger than $r_p$ to account for the redshift-space distortions created by peculiar velocities of multiplet members. We then find multiplets within these matches using a Union Find algorithm to identify connected components within the graph of galaxy pairs \citep{galler_improved_1964}. We set no maximum for the number of multiplet members. This is similar to the friends-of-friends algorithm used for identifying haloes in N-body simulations and for constructing group catalogues \citep{davis_evolution_1985, eke_galaxy_2004, robotham_galaxy_2011}. Note that unlike these catalogues, our goal is not to identify complete, gravitationally bound objects. We expect even nonviralized objects to contribute to our final measurement and so set no additional criteria such as completeness or velocity dispersion.

To explore the effectiveness of this algorithm to identify distinct multiplets, we created a catalogue of isolated multiplets, consisting only of multiplets where each member was a minimum of 2$r_p$ and 2$r_\parallel$ away from the nearest non-multiplet member. This had no significant effect on final results. We tested multiplets constructed from varying criteria, between $0.5 h^{-1}\text{Mpc}<r_p<1.0h^{-1} \text{Mpc}$ and $6.0 h^{-1}\text{Mpc}<r_\parallel<12h^{-1} \text{Mpc}$. We found no significant effect on the amplitude of the final signal when varying these parameters, so we selected cuts to maximise the signal-to-noise ratio (SNR) for the sample we model in Section~\ref{sec:interp}, LRGs. For all samples, we use $r_p = 1.0h^{-1}\text{Mpc}$ and $r_\parallel = 6.0h^{-1}\text{Mpc}$. 
Alternative multiplet definitions, such as scale cuts which depend on density or removing very close pairs (see Section~\ref{sec:sims}) may improve the SNR and are worth exploring in future works.

Properties of the DESI samples we identified multiplets in are shown in Table \ref{tab:samples} and demographics of the multiplets are shown in Fig.~\ref{fig:demographics}. This displays the number of members within each multiplet, which is most often two. It also shows the spatial size of each multiplet, determined by taking the maximum 3D distance between a multiplet member and the multiplet's centre in redshift space. Note that this can be greater than $r_p$ and $r_\parallel$ since we only limit the distance between multiplet members, not its overall size.

\subsection{Estimator formalism}\label{sec:formalism}

For each multiplet, we determine its projected orientation based on the 2D positions of its member galaxies in the plane of the sky. This orientation is then correlated with the positions of galaxies in a tracer sample. See Fig.~\ref{fig:schematic} for a schematic of the variables used.

Multiplet orientation $\Theta$ is determined by averaging the complex positions of its $N$ members relative to its centre. This centre is defined as the geometric average of member 2D positions. $\Theta$ is the angle corresponding to this average complex number:
\begin{equation}\label{eq:rel_ang}
    \begin{split}
    \epsilon_{\rm multiplet} = \frac{1}{N}\sum\limits_{i=1}^N r_i \exp{2i\theta_i} = a + bi \\
    \Theta = \frac{1}{2}\arctan{\frac{b}{a}}
    \end{split}
\end{equation}
For each member $i$, $r_i$ is the projected distance to the multiplet centre and $\theta_i$ is its projected position angle relative to the centre. We do not consider the full ellipticity of the multiplet, i.e. axis ratio, because this is meaningless for multiplets with two members and it is not expected to increase our signal-to-noise ratio. For single galaxies, measurements with the SDSS-III BOSS LOWZ sample have found that this can provide comparable constraints \citep{singh_intrinsic_2015}. We find a similar result with DESI Year 1 LRGs and Legacy Imaging (Fig.~\ref{fig:imaging_comparison}).

This orientation angle is then measured relative to the tracer sample. In most cases, the tracer sample is the same as the one used to identify multiplets. For each pair, consisting of a galaxy multiplet and a tracer, the relative angle is the difference between the position angle of the multiplet relative to the tracer, $\phi$, and the multiplet's orientation:
\begin{equation}
    \Phi = \Theta - \phi
\end{equation}
The relevant quantity is $\cos({2\Phi})$, as function of the projected separation between the multiplet and the tracer, $R$. This is then averaged over every multiplet-tracer pair.
This is similar to conventions in intrinsic alignments and ensures the relative angle is invariant under rotation. Since we are mostly measuring the orientation of a pair of galaxy positions relative to a distant tracer, this estimator is similar to a squeezed three-point correlation function. For multiplets with more than two members, the pair of points describe an N-weighted orientation.\par
The code for these measurements, and other analysis throughout the paper, is available in the repository \href{https://github.com/cmlamman/spec-IA}{spec-IA}.

\begin{figure} 
\begin{center}
\includegraphics[width=.4\textwidth]{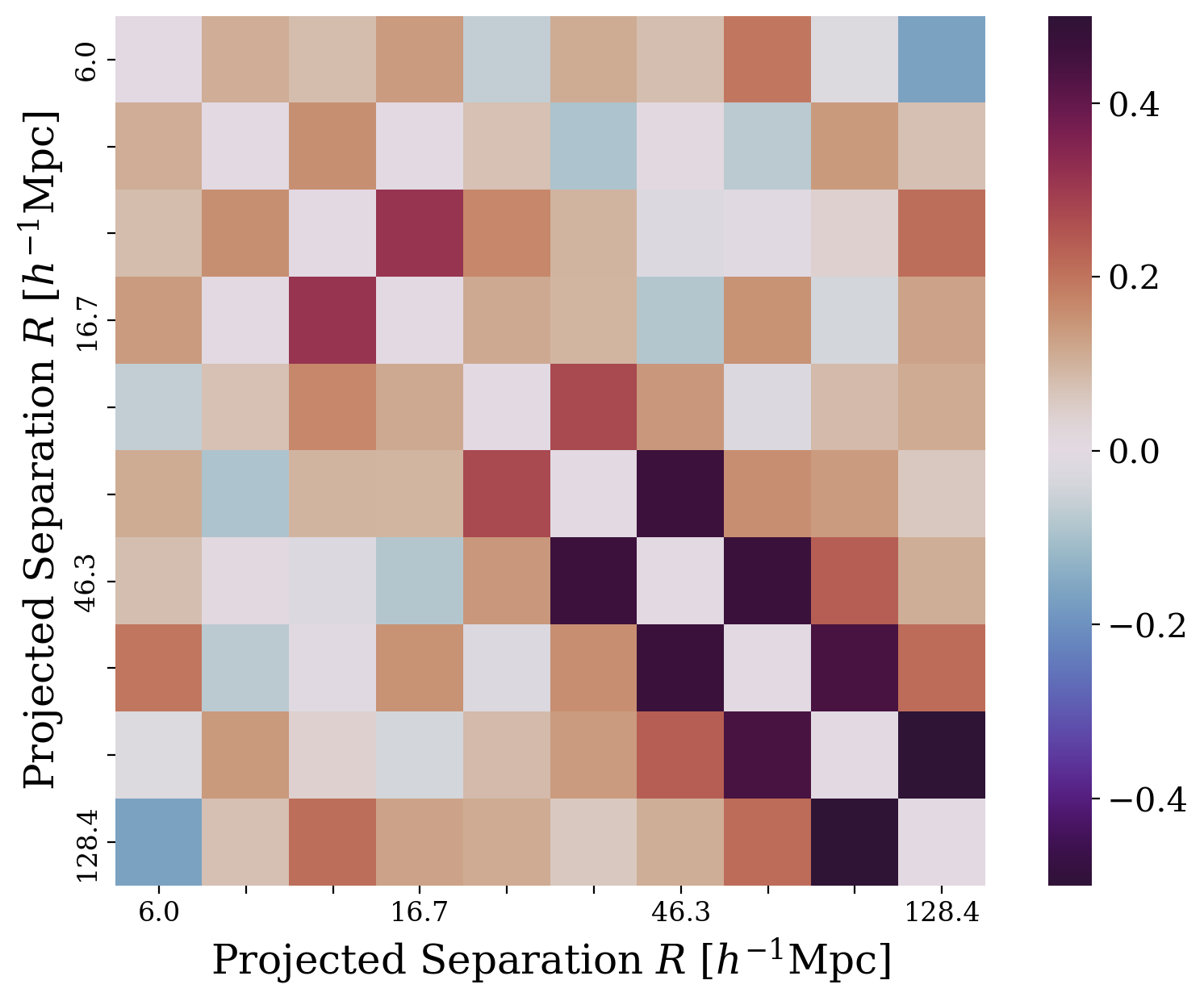}
\end{center}
\caption{The reduced covariance matrix corresponding to the LRG signal in Fig.~\ref{fig:tracer_signals}. The identity matrix has been subtracted from this plot. There is covariance in the multiplet alignment measurement between bins of projected separation, particularly in the largest bins.}
\label{fig:main_cov}
\end{figure}

\begin{figure} 
\includegraphics[width=.47\textwidth]{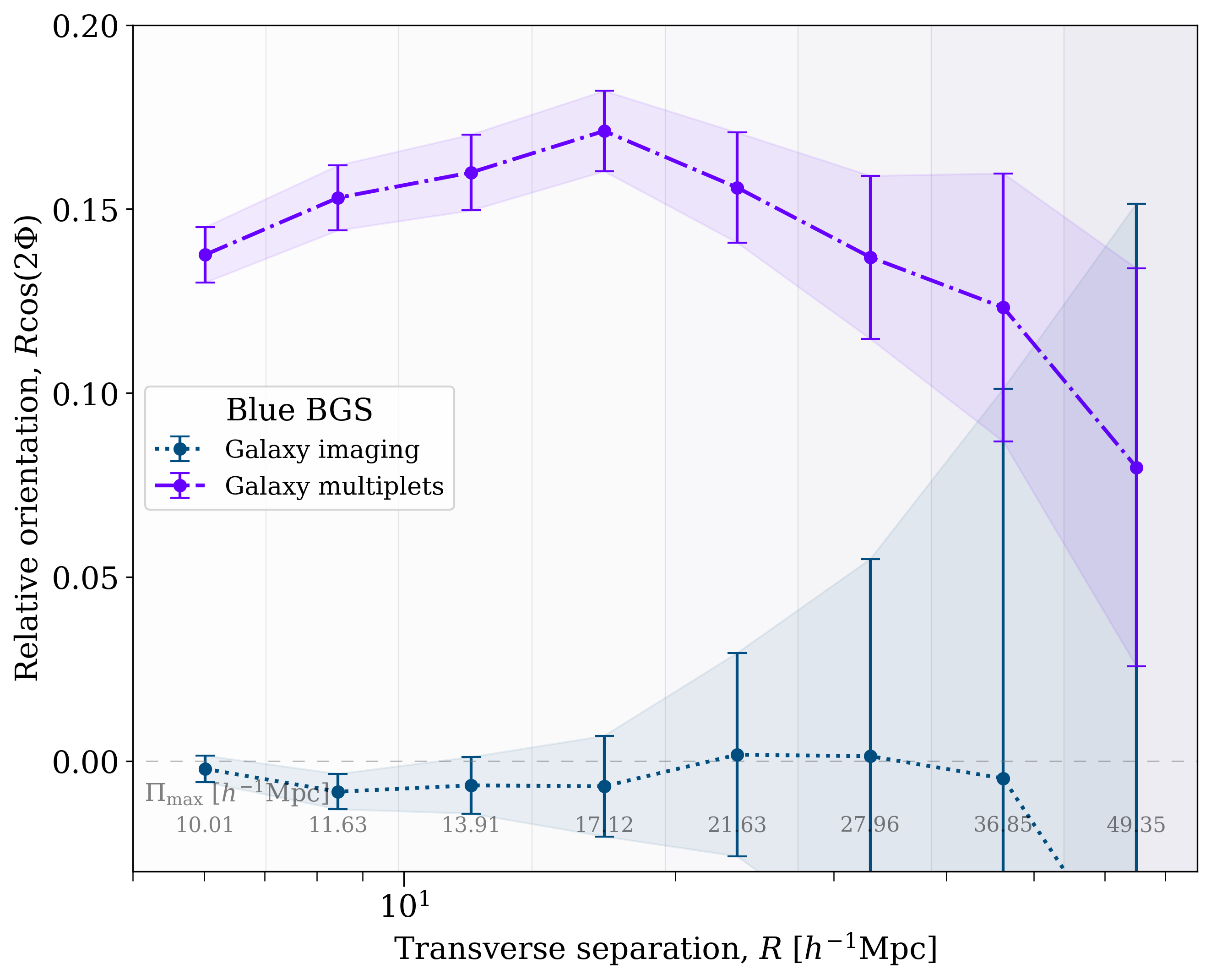}
\caption{A demonstration of the advantages of using multiplet alignment. Here we show the tidal alignment of galaxy and multiplet orientations within a dense, blue sample. The alignment of individual galaxies is highly sensitive to survey geometry and, as expected, consistent with zero. However, the alignment of multiplets displays a clear signal.}  
\label{fig:blue_bgs}
\end{figure}

\begin{figure*}
\centering
\begin{subfigure}[b]{0.47\textwidth}
    \includegraphics[width=\textwidth]{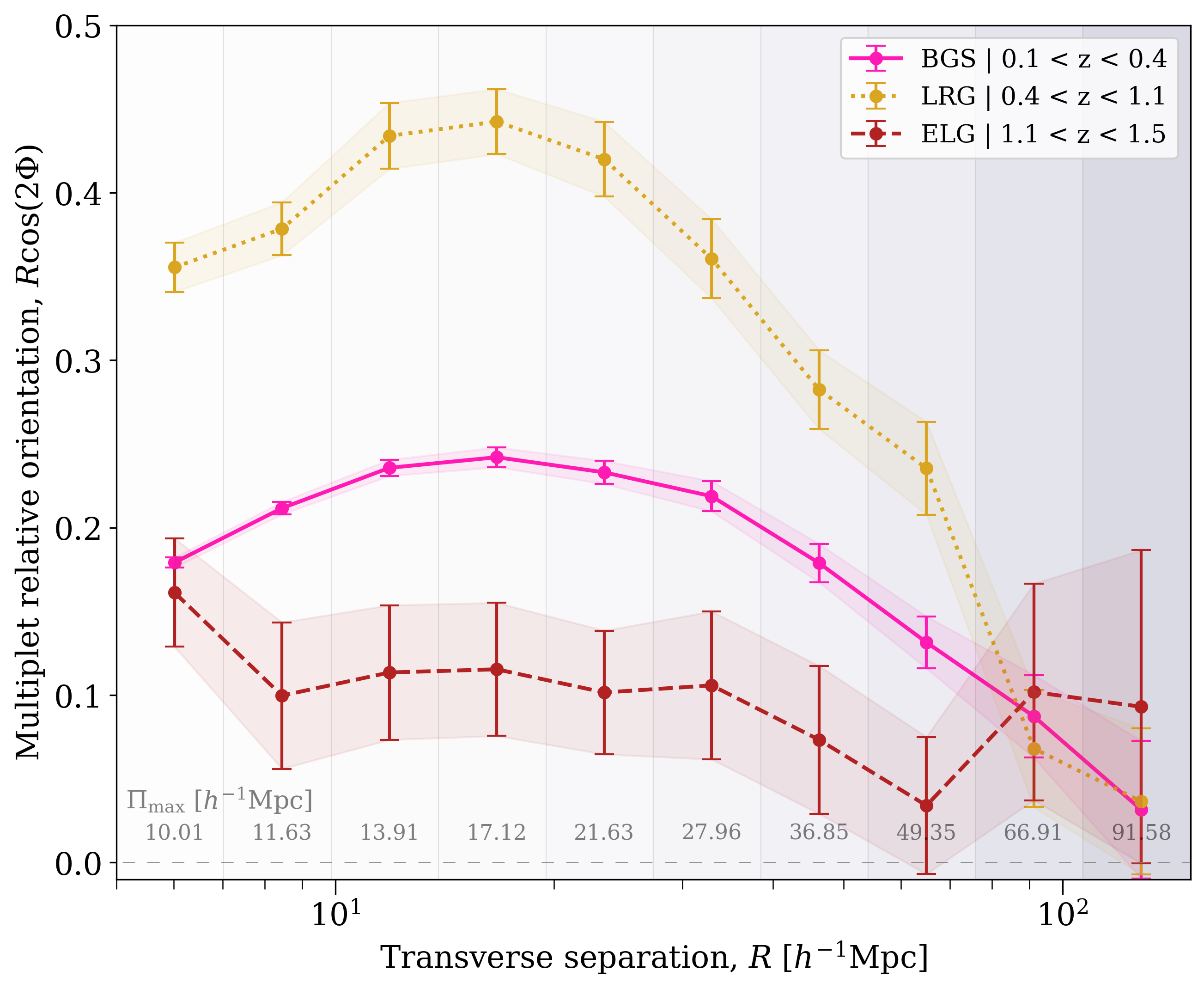}
    \caption{}
    \label{fig:tracer_signals}
\end{subfigure}
\hfill
\begin{subfigure}[b]{0.48\textwidth}
    \includegraphics[width=\textwidth]{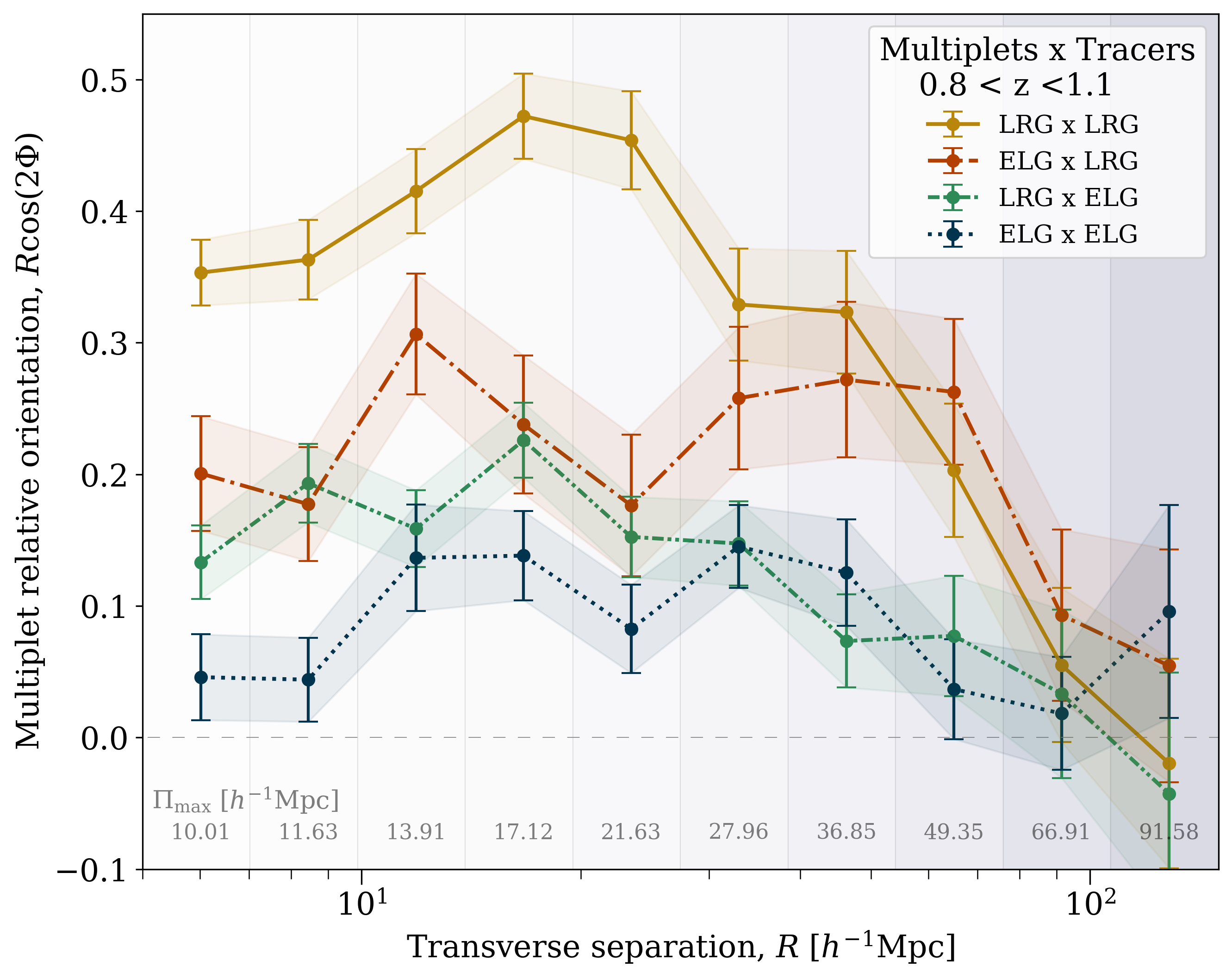}
    \caption{}
    \label{fig:ELGLRG_signals}
\end{subfigure}

\begin{subfigure}[b]{0.47\textwidth}
    \includegraphics[width=\textwidth]{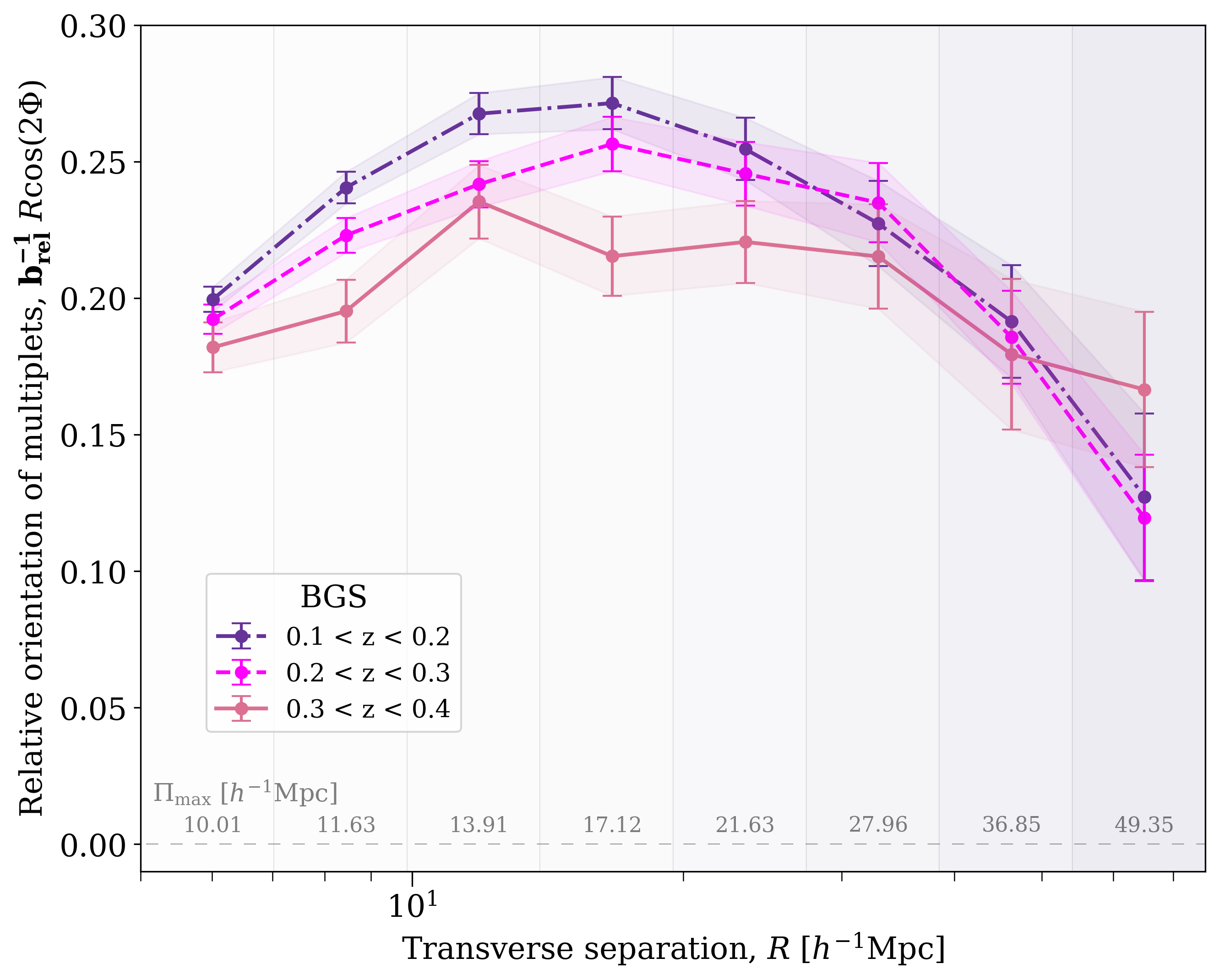}
    \caption{}
    \label{fig:bgs_signals}
\end{subfigure}
\hfill
\begin{subfigure}[b]{0.47\textwidth}
    \includegraphics[width=\textwidth]{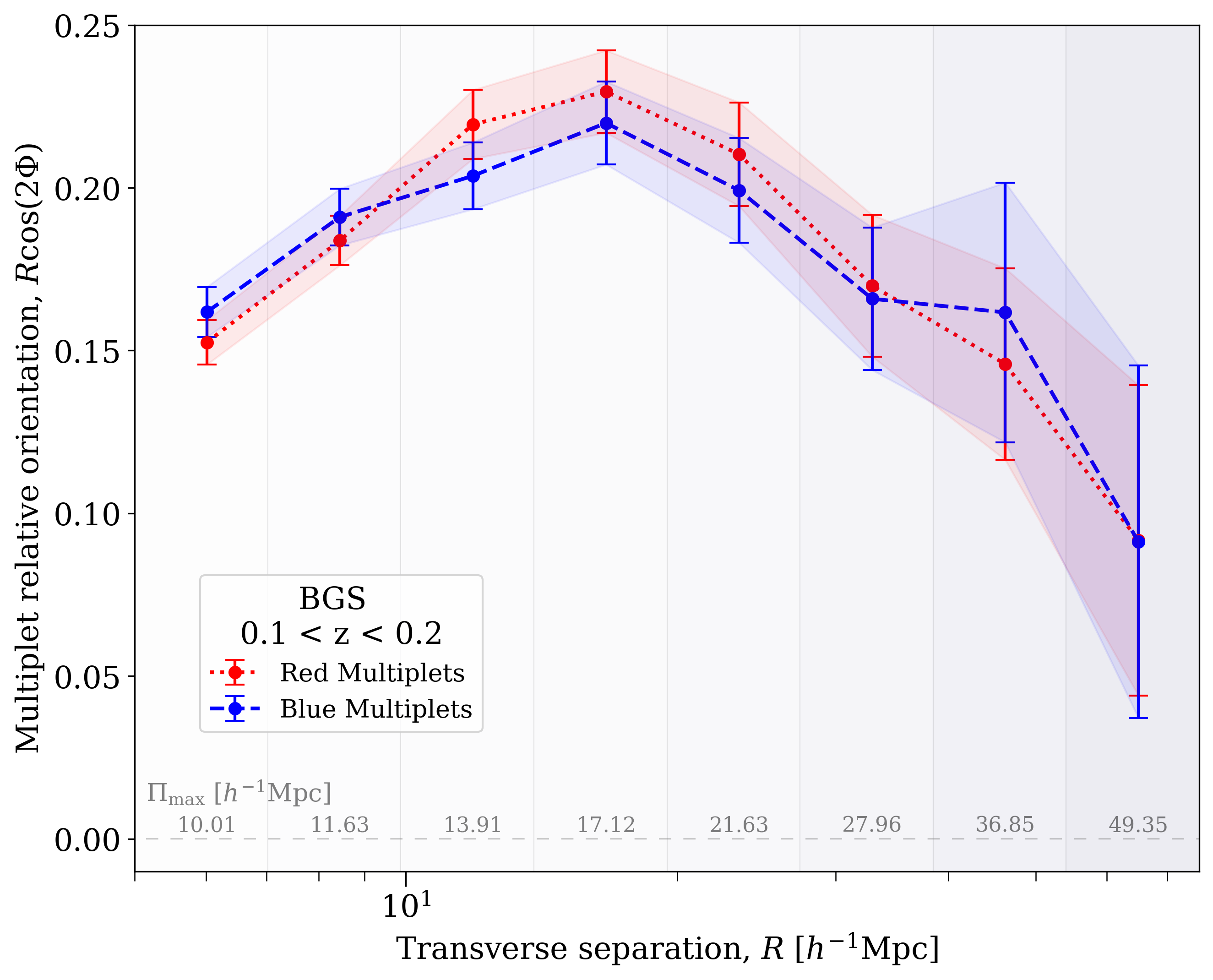}
    \caption{}
    \label{fig:blue_vs_red_bgs}
\end{subfigure}
\caption{Correlations between the projected orientations of galaxy multiplets and density for different galaxy samples as a function of projected separation, $R$. The measurement in each $R$ bin utilises a different value of $\Pi_{\rm max}$, indicated by the shaded regions and marked at the bottom of each plot. $\Pi_{\rm max}$ is the maximum line-of-sight distance between a multiplet-tracer pair. Unless otherwise indicated, each measurement uses the same catalogue for multiplet orientations and density tracers. (a) The signal for each tracer type, with no adjustments made for differences in clustering between samples. LRGs have the highest galaxy bias and their signal is the one we focus on reproducing in Section~\ref{sec:interp}. The signal is especially clear for the dense BGS sample. Although a sparse sample, we also detect a signal with ELGs beyond redshift 1. (b) explores cross-correlations between ELGs multiplets, ELG tracers, LRGs multiplets, and LRG tracers in their overlapping region, $0.8<z<1.1$. Based on the comparison in (d), we expect similar scale-dependence of these alignments. Redshift subsets of the BGS sample are shown in (c). Here we account for differences in the galaxy bias and its evolution by scaling each measurement relative to the bias in the middle redshift bin. From lowest to highest redshift bin, the rescaling factors are 1.12, 1.0, and 0.80. (d) displays the alignment of multiplets in red and blue subsamples of the lowest redshift BGS galaxies, relative to the full BGS sample. We find no obvious difference in scale dependence, demonstrating the potential of utilising blue galaxies to trace the tidal field similarly to red galaxies.}
\label{fig:all_signals}
\end{figure*} 

\subsection{Measurement}\label{sec:measurment}

When measuring the projected orientation of multiplets relative to a tracer catalogue, we limit the multiplet-tracer pairs to a line-of-sight separation that is unique to each bin of projected separation, $\Pi_{\rm max}(R_{\rm bin})$. This is to maximise the signal-to-noise of our measurement. In the case of positive tidal alignment, shapes are elongated along the stretching direction of the tidal field. In this situation, the tidal field along the line of sight will not induce a measurable orientation in the plane of the sky. Therefore, multiplet-tracer pairs that are close in the plane of the sky but distant along the line-of-sight direction will have a relatively low contribution to the total alignment signal. At larger projected separations there is more contribution from radially distant galaxies and it becomes advantageous to increase $\Pi_{\rm max}$. We chose $\Pi_{\rm max}(R_{\rm bin}) = 6h^{-1}\text{Mpc} + \frac{2}{3}R_{\rm bin}$ based on the signal-to-noise ratio of our final LRG signal. Our model estimate is computed in these same $R$ bins. Throughout plots in this paper, the varying values of $\Pi_{\rm max}$ are shown through shaded regions and marked explicitly. We use this projected statistic, as opposed to keeping the measurement as a function of $r_p$ and $r_\|$, because most of the signal is along the LOS for tidal alignments due to the projection of shapes. Additionally, a projected statistic allows for more direct modelling as it is less sensitive to redshift-space distortions (Figure \ref{fig:mock_rsd}). 

For each measurement, we separate the multiplet catalogue into 100 sky regions by right ascention and declination, with equal numbers of multiplets in each. The orientation of multiplets are measured separately in each region, but relative to the full tracer sample. Our final measurement is the mean and standard error of these 100 measurements. For the densest of our samples, BGS, we use 144 regions and adopt a more memory-conscious binning strategy. In this instance we compute the average signal in each $R_{\rm bin}$ before averaging over the sky regions. This marginally increases the measurement noise but is more practical for samples with many multiplet-tracer matches. 

DESI's Year one survey has an irregular footprint with varying levels of survey completeness. For a plot of the Year one LRG and BGS footprint, see Figure 4 of \citealt{krolewski_impact_2024}; more details on completeness variation can be found in \citet{DESI2024.II.KP3}. We find the signal to be sensitive to survey geometry on large scales. To account for this, for every measurement we also measure the orientation of galaxy multiplets relative to random catalogues designed to match DESI's Y1 footprint. The average of measurements with multiple random catalogues is subtracted from the initial measurement. Across samples, we see a turnover in the multiplet-random signal around 80 $h^{-1}$Mpc. We see no evidence of anisotropy in the orientations of multiplets, so this systematic "tangential alignment" at large separations is likely to be due to the footprint of the tracers, which spans a narrow band in right ascension. This pattern is not present when measuring the signal in isolated square regions.

The multiplet alignment measurements can be found in Fig.~\ref{fig:all_signals}, and the covariance matrix for the LRG multiplet alignment in \ref{fig:main_cov}. Within the four bins we used to scale our model in Section~\ref{sec:interp}, between 20 - 70 $h^{-1}$ Mpc, the LRG signal has a detection significance of 15.8$\sigma$. We find a detection of alignment out to $100 h^{-1}$Mpc in all samples, including the highest-redshift ELG bin at $1.1<z<1.5$. This is shown in Fig.~\ref{fig:tracer_signals}, along with the full LRG and BGS samples. Here we have made no adjustment for clustering differences between samples; this plot demonstrates the alignment strength and scale dependence of each tracer, not any redshift dependence. For each measurement, we use the same galaxies to construct multiplets and the tracer catalogue. The exception is the overlapping LRG and ELG region of $0.8<z<1.1$, where we measure each cross-correlation between the samples (Fig.~\ref{fig:ELGLRG_signals}). Although we expect each of these signals to display a similar scale dependence, it is difficult to assess with this sparse sample. Therefore we examine the red-blue dependence within the densest of our catalogues, BGS.

We split the BGS catalogue into redshift subsamples (Fig.~\ref{fig:bgs_signals}). In this plot we account for the galaxy bias and its evolution across redshifts. Intrinsic properties of BGS also vary across redshift, so this plot should not be interpreted as a redshift evolution. For instance, the highest redshift bins contains the most luminous galaxies, which are known to display higher alignment. Despite this, we still see a high signal from the lower-mass, low-redshift galaxies. Galaxies in the lowest BGS redshift bin have stellar masses of around $10^{10}M_\odot$ \citep{hahn_desi_2023}. Fig.~\ref{fig:blue_vs_red_bgs} shows the multiplet alignment for blue vs red galaxies in this same sample. The colour cuts selection of these subsamples are described in Section~\ref{sec:desi_catalogue} and their alignments were measured relative to the full BGS catalogue. The red and blue samples display similar amplitudes and scale dependence. This shows that blue galaxies can be used to trace the tidal field similarly to red galaxies, which is a promising result for measuring alignments beyond redshift 1.

Compared to the alignment of individual galaxies, we expect the alignment of galaxy multiplets to be well-suited to samples that are especially dense and samples of blue galaxies. To directly demonstrate this, within the BGS Blue sample we measure the intrinsic alignment of individual galaxies using imaging from the Legacy Imaging Survey. Here we find a strong dependence on survey geometry. This is accounted for through the randoms, but results in a large statistical error at large separations. As expected, the alignment of these faint blue galaxies is consistent with zero at all separations. However, galaxy multiplets in the same catalogue display a clear alignment signal (Fig.~\ref{fig:blue_bgs}).

\begin{figure} 
\includegraphics[width=.48\textwidth]{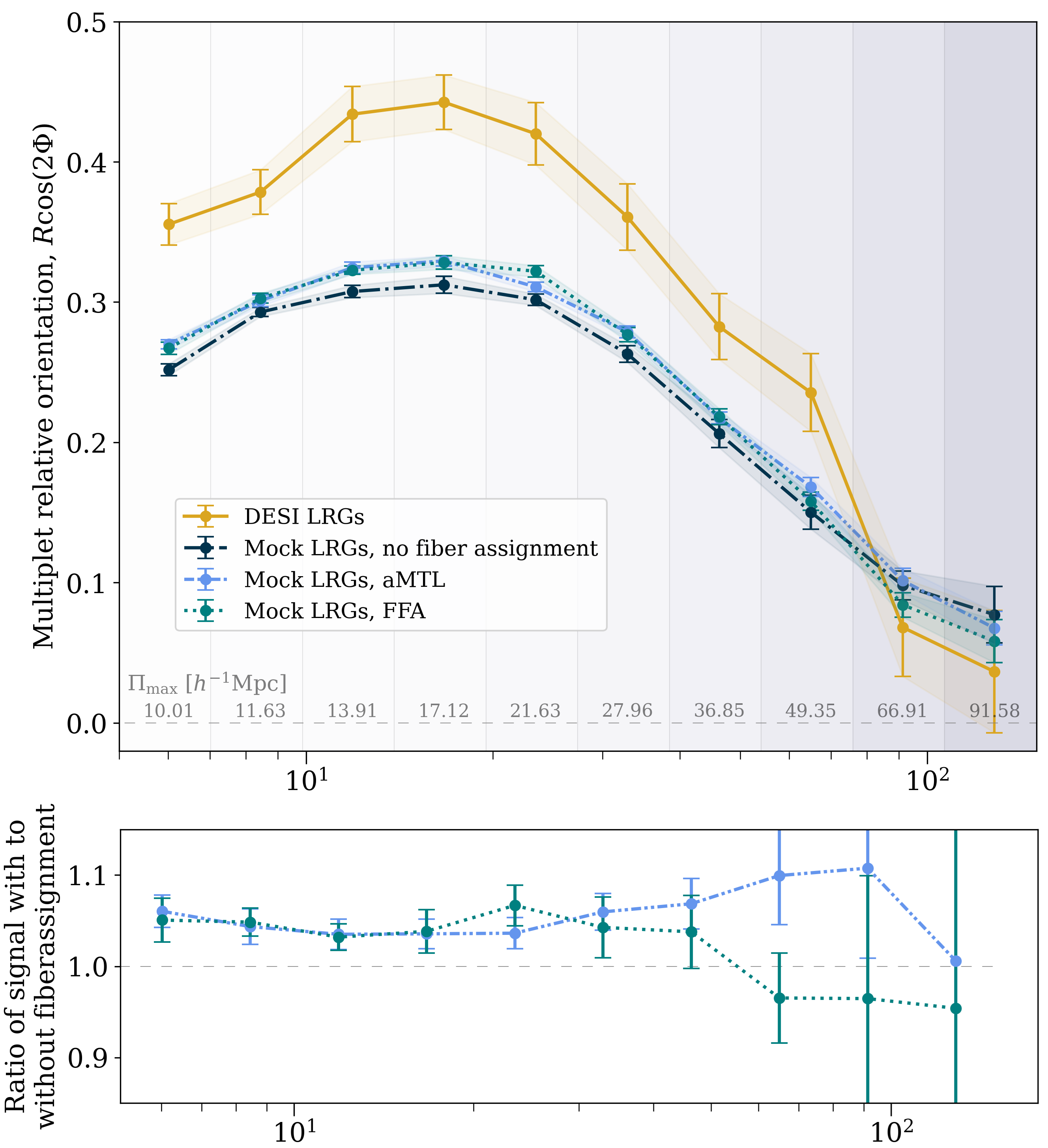}
\caption{A comparison between DESI Year 1 LRGs and several LRG mock catalogues. The mock catalogues contain different implementations of fiber assignment: None, aMTL, and FFA (as described in Section~\ref{sec:desi_catalogue}). The ratio of the mock signals with fiber assignment to that without is shown in the bottom panel. Fiber assignment causes a 4-5\% enhancement of the signal.}
\label{fig:mock_signal}
\end{figure}

\begin{figure} 
\includegraphics[width=.48\textwidth]{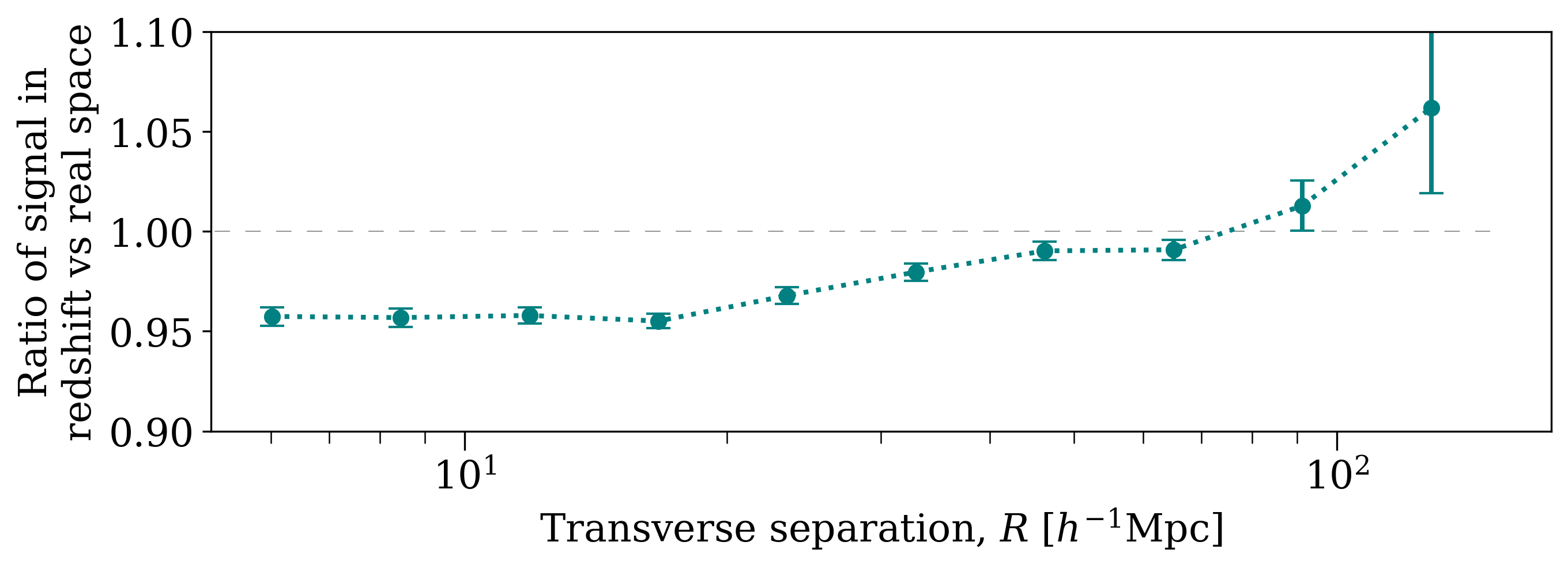}
\caption{This is an assessment of the impact of RSD on the multiplet alignment signal. Here we plot the ratio between the aMTL signal in Fig.~\ref{fig:mock_signal} and a version where the shape-tracer correlations were measured in real space. The two measurements differ by about 5\% on these scales.}
\label{fig:mock_rsd}
\end{figure}

\section{Interpretation}\label{sec:interp}
In this Section~we explore the modelling of multiplet alignments through simulations and theory, using the LRG sample as a case study. This is because LRGs have a large galaxy bias, they display clear alignment of both individual and multiplet orientations, and there exist associated DESI mocks that are designed for reproducing measurements of large-scale structure.

\subsection{Comparison to simulations}\label{sec:sims}
We reproduce our measurements with three versions of DESI’s Y1 LRG mock catalogues: one without fiber assignment and two with different implementations of fiber assignment, aMTL and FFA, as described in Section~\ref{sec:desi_catalogue}. These fiber assignment catalogues and their weights are designed to reproduce 2-point clustering statistics. The average measurement of 25 simulations for each mock catalogue and their standard error compared to the true LRG signal is shown in Fig~ \ref{fig:mock_signal}. We find no significant difference in the number of multiplets found, but they display lower multiplet alignment. This is probably a reflection of the underlying simulation’s inability to capture higher-order clustering effects. However, they sufficiently reproduce the shape of the signal.

aMTL is the most realistic simulation of fiber assignment, but we do not find a significant difference between the two mock catalogues which include fiber assignment. It is interesting to note that the signal is marginally higher for the fiber assignment catalogues, which can be seen in the lower panel of Fig.~\ref{fig:mock_signal}. This is probably because galaxies very close to a multiplet's centre are more affected by nonlinear dynamics and therefore dilute large-scale correlations. Individual galaxy shapes display higher alignment in their outer regions for the same reason \citep{singh_intrinsic_2016, georgiou_dependence_2019}. Fiber assignment under-selects close pairs, effectively removing some of this dilution. To test this, we limited the projected separation of the initial pairs used to make the LRG multiplet catalogue to $r_p > 0.5h^{-1}$Mpc and find a similar enhancement of the signal, 10\% between 6 -- 60 $h^{-1}$Mpc. This may be a useful addition to future studies of multiplet alignment.

We do not include the effects of RSD in our analytic model (Section~\ref{sec:interp}), so to test this assumption we reproduce the aMTL measurement in real space. Here, galaxy multiples are still found in redshift space, but the multiplet-tracer correlations are measured using the true positions of the multiplet centres and tracers. The effects of RSD on the tracer catalogue appear to make a 0 -- 5\% difference on scales beyond 10 $h^{-1}$Mpc (Fig.~\ref{fig:mock_rsd}).

\begin{figure} 
\includegraphics[width=.48\textwidth]{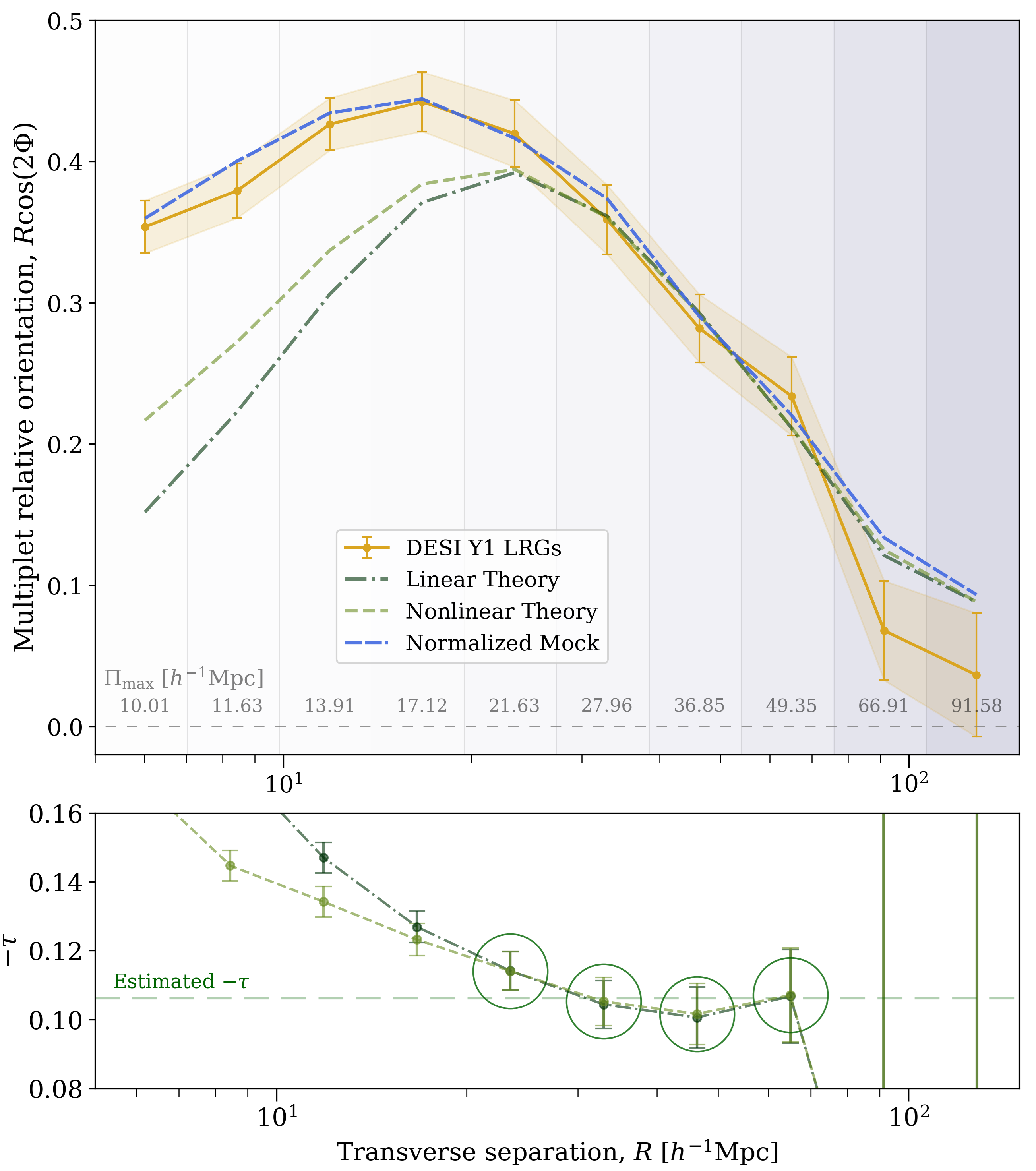}
\caption{Comparison between the LRG multiplet alignment and the model predictions using both a linear and nonlinear matter power spectrum. These model predictions have been normalized using a $\tau$ value estimated from the bin measurements circled in the bottom panel. We also show a normalized measurement of the multiplet alignment made with LRG aMTL mock catalogues (Section~\ref{sec:sims}), which sufficiently reproduces the signal shape on these scales.}
\label{fig:model_comparison}
\end{figure}
\subsection{Modeling}\label{sec:modeling}

To quantify the connection between multiplet orientation and the underlying matter distribution, we assume a linear relationship between shapes and the tidal field. This common approach for large-scale intrinsic alignment assumes the ellipticity of objects are linearly related to the gravitational potential, which is described by either the linear (LA) or nonlinear (NLA) matter power spectrum \citep{bridle_dark_2007, hirata_intrinsic_2007}. The latter is often used for individual LRGs down to projected separations of around $6h^{-1}$Mpc \citep{singh_intrinsic_2015}. The amplitude of this relation is then calibrated by measurements.\par
Following the convention in \citet{lamman_intrinsic_2023, lamman_redshift-dependent_2024}, we describe the traceless tidal tensor as
\begin{equation}\label{eq:tidal_tensor}
    T_{ij} = \partial_i \partial_j \phi - \frac{1}{3} \delta^{\rm K}_{ij} \nabla^2 \phi,
\end{equation}
where $\nabla^2 \phi\propto \delta$ is given by the Poisson equation, with $\delta$ being the fractional overdensity. Constants are absorbed into the unmodeled amplitude of the correlation. $\delta^{\rm K}$ is the Kronecker delta. In Fourier space this is expressed as 
\begin{equation}
    T_{ij}(\vec r) = \int \frac{d^3k}{(2\pi)^3} \left( \frac{k_i k_j - \frac{1}{3}\delta^{\rm K}_{ij}k^2}{k^2}\right) \tilde\delta_m(\vec k) e^{i\vec k\cdot\vec r},
\end{equation}
where we have use $\tilde{\delta}$ to indicate a variable in Fourier space. Our measured signal is a projected quantity, where we define $\hat{z}$ to be along the line-of-sight. Therefore, for a projection with $\alpha$, $\beta$ = $\{x,y\}$ and using the relation $T_{xx}+T_{yy} = -T_{zz}$, the relevant projection of the tidal field is $(T_{\alpha\beta} + T_{zz}/2)$. 

In this study we characterize the relevant ``shapes'' of objects solely by orientation, instead of the full ellipticity. This axis-ratio component of shapes affects the amplitude of the signal as does any systematic misalignment of galaxy multiplets to the large-scale field caused by local dynamics. 
Our focus in this work is to explore how multiplet alignment traces the tidal field across large scales, without making any assumptions about the effect's amplitude. Therefore, we fold in the full ellipticity information and any misalignment effects into the signal amplitude, assuming neither display scale dependence at large separations. This is similar to the ``stick model'' employed for describing the positions and alignments of galaxies within haloes \citep{fortuna_halo_2021, schneider_halo_2010}. 

The projected ellipticity of galaxy multiplets can be described by the traceless tensor
\begin{equation}
    \epsilon_{\alpha,\beta} = \tau(T_{\alpha\beta} + \frac{1}{2}T_{zz}).
\end{equation}
$\epsilon_{\alpha,\beta}$ quantify the relative orientation of galaxy multiplets, as defined in Section~\ref{sec:formalism}, and $\tau$ is a parameterization of the shape's response to the tidal field. $\tau$ includes any effects from the full-shape information and any misalignment of shapes relative to the tidal direction.
The full complex ellipticity is described as
\begin{equation}\label{eq:E_complex}
    \epsilon = \tau [T_{xx} - T_{yy} + 2iT_{xy}].
\end{equation}

The quantity of interest is the expectation value of the cross-correlation between projected shapes and the matter field, $Q$: 
\begin{equation}
    \mathcal{E}_{\rm model}
    = \frac{1}{2}\langle \epsilon^* Q + \epsilon Q^* \rangle . 
\end{equation}
We describe the 3D matter field in a particular bin of transverse separation $R_{\rm bin}$ and line-of-sight separation $\pm\Pi_{\rm max}$ as
\begin{equation}\label{eq:Q}
    Q(R_{\rm bin}, \pm\Pi_{\rm max}) = \frac{\int d^3r W(\bar{r})\delta_g e^{2i\theta_r}}{\int d^3r W(\bar{r})(1 + \xi_{\epsilon g})} .
\end{equation}
Here, $\delta_g$ is the fractional matter overdensity, $\xi_{\epsilon g}$ is the shape orientation -- galaxy correlation function, $r$ is the 3D separation, and $\theta_r$ is the 3D relative angle. $W(\bar{r})$ is a function representing the bin selection, both an annulus in $R$ and $\pm\Pi_{\rm max}$ along the line of sight, i.e. $\hat{z}$. $\bar{r}$ is used to denote a binned quantity. The expansion of $\epsilon^* Q + \epsilon Q^*$ can be found in Appendix \ref{appendix:modeling}, and results in the expression
\begin{equation}\label{eq:E_model_0}
    \begin{split}
    \mathcal{E}_{\rm model} = \frac{-\tau}{\int d^3r W(\bar{r})(1 + \xi_{\epsilon g})}
    \int d^3r W(\bar{r}) \\
    \int \frac{dk_z}{\text{2}\pi} \int KdK J_2(KR)\frac{K^2}{k^2}P_{gm}(k)e^{ik_z z},
    \end{split}
\end{equation}

\noindent where $J_2$ is the second Bessel function of the first kind and $P_{gm}(k)$ is the galaxy-matter power spectrum. $k$ represented 3D position in Fourier space, $K$ represents the 2D position on the plane of the sky ($k_x$, $k_y$), and $k_z$ lies along the line of sight. $k^2 = K^2 + k_z^2$. 

The remainder of this Section~describes how we compute Equation \ref{eq:E_model_0}, by breaking it into the components we measure or calculate. Beginning with the denominator,
\begin{equation}
    \int d^3r W(\bar{r})(1 + \xi_{\epsilon g}) = \pi (R_{\rm max}^2 - R_{\rm min}^2)(2\Pi_{\rm max} + \bar{w}_p).
\end{equation}
$\bar{w}_p$ is the integrated 2-point cross-correlation function between the multiplet and tracer catalogue, $w_p(R)$, within an annulus of $R_{\rm min}$ and $R_{\rm max}$:
\begin{equation}
    \bar{w}_p(R_{\rm bin}) = \frac{1}{\pi (R_{\rm max}^2 - R_{\rm min}^2)} \int_{R_{\rm min}}^{R_{\rm max}} 2\pi R dR w_p(R).
\end{equation}
We further define $\mathcal{J}_2$, a binned version of the second Bessel function integrated over a given $R_{\rm bin}$:
\begin{equation}
    \mathcal{J}_2(K) = \frac{2}{(R_{\rm max}^2 - R_{\rm min}^2)} \int_{R_{\rm min}}^{R_{\rm max}} RdR J_2(KR),
\end{equation}
This can be solved analytically (Equation \ref{eq:fancyJ_solution}). Using the relation
\begin{equation}
    \frac{1}{2\Pi_{\rm max}} \int_{-\Pi_{\rm max}}^{\Pi_{\rm max}} dz e^{ik_z z} = \text{sinc}{(k_z\Pi_{\rm max})},
\end{equation}
we further define an expression of the relevant matter distribution for a given $\Pi_{\rm max}$:
\begin{equation}
\mathcal{P}_{\Pi}(K) = 2 \Pi_{\rm max} \int \frac{dk_z}{2\pi} \frac{K^2}{K^2 + k_z^2} P_{gm}\bigg(\sqrt{K^2 + k_z^2}\bigg) \text{sinc}(k_z\Pi_{\rm max}).
\end{equation}

\noindent In practice, for this we use the matter power spectrum and galaxy bias $b_{g}P_{mm}(k)$, with $b_{g}=1.99$ for DESI LRGs \citep{mena-fernandez_hod-dependent_2024}. Combining these expressions, the model prediction for our signal $\mathcal{E}(R)$ is simplified to
\begin{equation}
    \mathcal{E}_{\text{model}}(R) = \frac{-\tau}{(2\Pi_{\rm max} + \bar{w}_p)}\int KdK\mathcal{J}_2(K, R)\mathcal{P}_{\Pi}(K).
\end{equation}

We compute this numerically in bins of ($R_{\rm min}$, $R_{\rm max}$) with the corresponding $\Pi_{\rm max}$ value in each. The model prediction made with both a linear and nonlinear matter power spectrum can be seen in Fig.~\ref{fig:model_comparison}. The power spectra are from A\textsc{bacus}S\textsc{ummit} and evaluated at $z=0.8$. We normalize the models by taking their ratio to the large-scale signal, using the points circled in the lower panel of Figure  \ref{fig:model_comparison}. This results in an estimate $\tau$ for the LRG multiplets of $-0.106 \pm 0.002$ for both LA and NLA. We find that these models can sufficiently match the shape of our measurement only down to scales of 20 $h^{-1}$Mpc, while the LRG mock catalogue matches below 10 $h^{-1}$Mpc. The corresponding $\tau$ value for this mock is also $-0.106\pm0.002$. Therefore the NLA model is sufficient for very large scales, but fails to capture the non-linear dynamics between multiplets and tracers in the way that an N-body simulation can.

\begin{figure} 
\begin{centering}
\includegraphics[width=.45\textwidth]{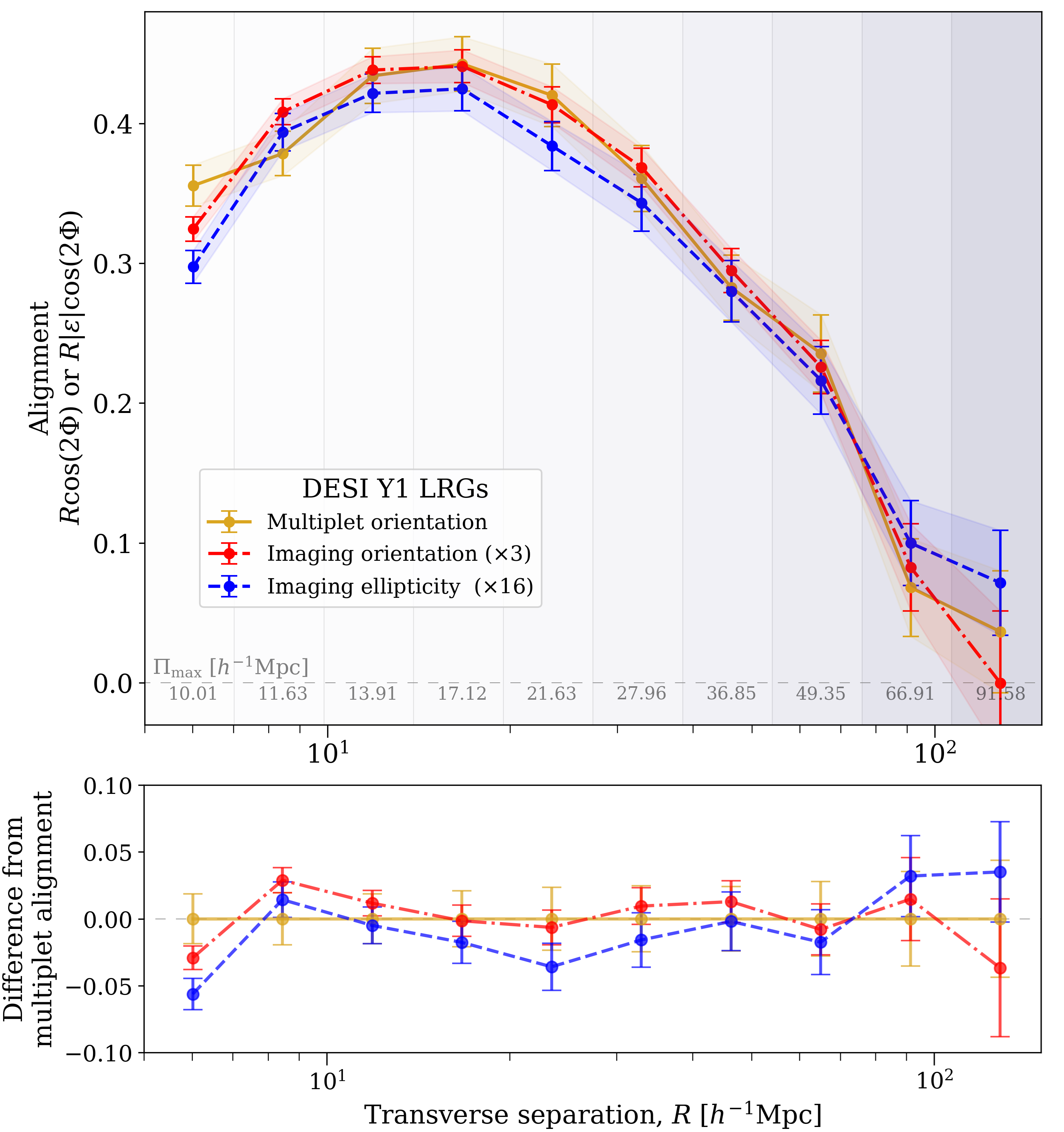}
\caption{Here we compare the alignment of galaxy multiplets to the alignment of individual galaxies within the DESI LRG sample. The yellow line shows the alignment of the orientations of LRG multiplets relative to positions of full sample (Section~\ref{sec:measurment}). The red line is the orientations of individual LRGs relative to the full sample, multiplied by 3 for an easier comparison. These two measurements model shapes as ``sticks'', described only by orientation. The blue line is the full shape alignment of LRGs, taking into account galaxy axis ratios and multiplied by 16 comparison. The bottom panel shows the difference in the points plotted above, highlighting the similar scale-dependence of each estimator. The average signal-to-noise for each of these measurements shown is 9.2 for multiplets, 15.4 for imaging, and 11.0 for imaging with ellipticity. These measurements were made with 105 thousand multiplets and 2.2 million individuals LRGs. \vspace{.1in}}
\label{fig:imaging_comparison}
\end{centering}
\end{figure}

The alignment amplitude is often characterized with $A_{\rm IA}$ \citep{catelan_correlations_2001, hirata_intrinsic_2004, blazek_tidal_2015}. $A_{\rm IA}$ describes the relationship between intrinsic galaxy shear, $\gamma^I_{ij}$, with the tidal tensor, $T_{ij}$, as defined in Eq. \ref{eq:tidal_tensor}. In the case of ``early alignment'', it is assumed that shapes are aligned at time of formation and then evolve with the matter field.
\begin{equation}
    \gamma^I_{ij} = -A_{\rm IA}(z)C_1\frac{\rho_{\text{m},0}}{D(z)}T_{ij}
\end{equation}
Here, $\rho_{\text{m},0}$ is the matter density, $D(z)$ is the growth factor, normalized so $\bar{D}(z)  =(1+z)D(z)$ is unity at matter domination, and $C_1$ is a historical normalization constant of $5\times 10^{-14} M_\odot^{-1}h^{-2}\text{Mpc}^3$ \citep{brown_measurement_2002}. The relationship to our alignment amplitude $\tau$ is
\begin{equation}
    A_\mathrm{IA} (z) = -\frac{\tau}{C_1} \frac{D(z)}{\rho_{\rm m,0}}
\end{equation}
 For our ``stick'' model of LRG multiples, this corresponds to an average value of $A_{\rm IA}= 5.7\pm0.1$. For reference, the corresponding stick alignment of the same sample using individual galaxies and Legacy Survey Imaging is $A_{IA} = 1.96\pm0.001$, about 5 times higher than when using the full-shape information (Fig.~\ref{fig:imaging_comparison}). 
 For this measurement we use the ellipticity definition
 \begin{equation}
     \epsilon_+ = \frac{a-b}{a+b}\cos{2\theta},
 \end{equation}
 based on the galaxy major and minor axis, $a$ and $b$, and orientation, $\theta$. 
 
 Fig.~\ref{fig:imaging_comparison} is also a useful demonstration of how, although very different amplitudes, the alignment of multiplet orientation has the same scale dependence of shape alignment and can be modeled similarly. Additionally, mutliplet alignment produces a comparable signal-to-noise measurement as full shape alignment, with less than 5\% of the objects. While multiplet alignment does not necessarily outperform individual galaxies within the LRG sample, it is promising for denser regions or samples that show weaker intrinsic galaxy alignment.

\section{Conclusion}\label{sec:conclusion}
In this work we explore the potential of multiplet alignment for large spectroscopic surveys through DESI's Year 1 data. These multiplets mostly consist of 2-4 members within $1h^{-1}$Mpc of each other. By measuring their orientations relative to the galaxy-traced tidal field, we detect an intrinsic alignment signal out to projected separations of 100 $h^{-1}$Mpc and beyond redshift $1$. Advantages of this galaxy multiplet alignment over the alignment of individual galaxies depend on properties of the galaxy catalogue, including morphology, density, and imaging quality. We find similar signals regardless of galaxy colour or luminosity, which is a promising result for measuring the tidal field with galaxy populations that typically display little or no intrinsic shape alignment. 

 Using the LRG sample as a case study, we reproduce the LRG measurement with mock catalogues from the A\textsc{bacus}S\textsc{ummit} N-body simulations, finding they under-predict the signal amplitude but match its shape. Using a nonlinear tidal alignment model, we find an amplitude parameter $\tau=-0.106\pm0.002$, which characterizes the response of multiplet orientations to the tidal field. This modelling matches the measured signal above scales of 20$h^{-1}$Mpc but fails to capture nonlinear effects at smaller scales, unlike the N-body prediction.

The multiplet alignment signal could be improved by supplementing multiplet catalogues with imaging, by identifying additional galaxies close to spectroscopic targets. Additional improvements could be made by weighting the shapes of multiplets based on member luminosity, or weighting the alignment by multiplet richness. Although we focus on modelling LRGs for this estimator, they are not necessarily the most optimal application. The signal is especially clear for the dense BGS region and warrants further exploration into sub-trends within the population, such as redshift and luminosity dependence. 

Compared to measurements of galaxy shape alignments, the use of spectroscopically identified multiplets mitigates systematic effects from imaging and shape measurements, and can extend intrinsic alignment studies to samples that do not display intrinsic shape alignments (Fig.~\ref{fig:blue_bgs}). The remaining 4 years of the DESI survey will significantly increase the size and comoving density of the ELG sample, allowing for better measurements of intrinsic alignment at higher redshifts. 

Although we describe multiplets as distinct objects throughout this work and model their orientations with intrinsic alignment conventions, they are not necessarily virialized systems or a proxy for dark matter halo shapes. The multiplet alignment estimator is essentially a squeezed three-point correlation function: the orientation of a pair of galaxies relative to a distant tracer, except in several cases the orientation of the galaxy pair is determined by all nearby galaxies. Our measurement shows that the angular dependence of the squeezed limit is coupled to the linear tidal field. This indicates that there is a directionality of nonlinear collapse preserved even into the non-linear regime, at the scales of the multiplet sizes of around 1 $h^{-1}$Mpc.

In principle, this measurement can be used to produce an intrinsic shear map and reconstruct the underlying matter field. Unlike the shear measurements from weak lensing, intrinsic shear preserves line-of-sight information. However, as in shear measurements, there will be impacts from the foreground mass on the multiplet shape polarization and therefore any inference of the tidal field. An advantage of using galaxy multiples to trace the 3D mass field is that the estimator has systematics that are distinct from the galaxy field, and may provide an avenue to more precisely measure large-scale modes. The difficulty lies in determining the modelling amplitude, $\tau(z)$. This could potentially be determined through hydrodynamic simulations or by calibrating $\tau$ with a weak lensing map and an assumed amplitude of the redshift-dependent matter power spectra. Such a mass map may prove useful for certain applications, despite the underlying cosmological dependence. With the right sample and understanding of the modelling amplitude, this could be a unique way to explore the large-scale matter field in future surveys.

\section*{Acknowledgements}
The authors wish to acknowledge useful conversations with Jonathan Blazek, Elisa Chisari, Thomas Bakx, and Christos Georgiou at the LILAC workshop, hosted at the Center for Astrophysics | Harvard \& Smithsonian. They also thank the DESI internal reviewers, Carolina Cuesta-Lazaro and Jiamin Hou for feedback on the paper.

This material is based upon work supported by the U.S.\ Department of Energy under grant DE-SC0013718, NASA under ROSES grant 12-EUCLID12-0004, and the Simons Foundation.

This material is based upon work supported by the U.S. Department of Energy (DOE), Office of Science, Office of High-Energy Physics, under Contract No. DE–AC02–05CH11231, and by the National Energy Research Scientific Computing Center, a DOE Office of Science User Facility under the same contract. Additional support for DESI was provided by the U.S. National Science Foundation (NSF), Division of Astronomical Sciences under Contract No. AST-0950945 to the NSF’s National Optical-Infrared Astronomy Research Laboratory; the Science and Technology Facilities Council of the United Kingdom; the Gordon and Betty Moore Foundation; the Heising-Simons Foundation; the French Alternative Energies and Atomic Energy Commission (CEA); the National Council of Science and Technology of Mexico (CONACYT); the Ministry of Science and Innovation of Spain (MICINN), and by the DESI Member Institutions: \url{https://www.desi.lbl.gov/collaborating-institutions}.

The DESI Legacy Imaging Surveys consist of three individual and complementary projects: the Dark Energy Camera Legacy Survey (DECaLS), the Beijing-Arizona Sky Survey (BASS), and the Mayall $z$-band Legacy Survey (MzLS). DECaLS, BASS and MzLS together include data obtained, respectively, at the Blanco telescope, Cerro Tololo Inter-American Observatory, NSF’s NOIRLab; the Bok telescope, Steward Observatory, University of Arizona; and the Mayall telescope, Kitt Peak National Observatory, NOIRLab. NOIRLab is operated by the Association of Universities for Research in Astronomy (AURA) under a cooperative agreement with the National Science Foundation. Pipeline processing and analyses of the data were supported by NOIRLab and the Lawrence Berkeley National Laboratory. Legacy Surveys also uses data products from the Near-Earth Object Wide-field Infrared Survey Explorer (NEOWISE), a project of the Jet Propulsion Laboratory/California Institute of Technology, funded by the National Aeronautics and Space Administration. Legacy Surveys was supported by: the Director, Office of Science, Office of High Energy Physics of the U.S. Department of Energy; the National Energy Research Scientific Computing Center, a DOE Office of Science User Facility; the U.S. National Science Foundation, Division of Astronomical Sciences; the National Astronomical Observatories of China, the Chinese Academy of Sciences and the Chinese National Natural Science Foundation. LBNL is managed by the Regents of the University of California under contract to the U.S. Department of Energy. The complete acknowledgments can be found at \url{https://www.legacysurvey.org/}.

Any opinions, findings, and conclusions or recommendations expressed in this material are those of the author(s) and do not necessarily reflect the views of the U. S. National Science Foundation, the U. S. Department of Energy, or any of the listed funding agencies.

The authors are honored to be permitted to conduct scientific research on Iolkam Du’ag (Kitt Peak), a mountain with particular significance to the Tohono O’odham Nation.

\section*{Data Availability}
The DESI Legacy Imaging Survey is publicly available at \href{https://www.legacysurvey.org/}{legacysurvey.org} and DESI's Early Data Release is  available at \href{https://data.desi.lbl.gov/doc/releases/edr/}{data.desi.lbl.gov/doc/releases/edr/}. Iron covers the DESI Year 1 sample and will be  released as part of DESI Data Release 1 (DR1) \citep{DESI2024.I.DR1}. A\textsc{bacus}S\textsc{ummit} simulations are available at \href{https://abacusnbody.org/}{abacusnbody.org}. \par
Data plotted in this paper can be downloaded from \href{https://zenodo.org/records/13230864}{zenodo.org/records/13230864}.


\bibliographystyle{mnras}
\bibliography{references} 

\vspace{.2in}
\noindent \textbf{Author Affiliations} \\

\noindent\textit{$^{1}$Center for Astrophysics $|$ Harvard \& Smithsonian, 60 Garden Street, Cambridge, MA 02138, USA\\
$^{2}$Departamento de F\'isica, Universidad de los Andes, Cra. 1 No. 18A-10, Edificio Ip, CP 111711, Bogot\'a, Colombia\\
$^{3}$Observatorio Astron\'omico, Universidad de los Andes, Cra. 1 No. 18A-10, Edificio H, CP 111711 Bogot\'a, Colombia\\
$^{4}$Lawrence Berkeley National Laboratory, 1 Cyclotron Road, Berkeley, CA 94720, USA\\
$^{5}$Physics Dept., Boston University, 590 Commonwealth Avenue, Boston, MA 02215, USA\\
$^{6}$Dipartimento di Fisica ``Aldo Pontremoli'', Universit\`a degli Studi di Milano, Via Celoria 16, I-20133 Milano, Italy\\
$^{7}$Department of Physics \& Astronomy, University College London, Gower Street, London, WC1E 6BT, UK\\
$^{8}$Instituto de F\'{\i}sica, Universidad Nacional Aut\'{o}noma de M\'{e}xico,  Cd. de M\'{e}xico  C.P. 04510,  M\'{e}xico\\
$^{9}$University of California, Berkeley, 110 Sproul Hall \#5800 Berkeley, CA 94720, USA\\
$^{10}$Institut de F\'{i}sica d'Altes Energies (IFAE), The Barcelona Institute of Science and Technology, Campus UAB, 08193 Bellaterra Barcelona, Spain\\
$^{11}$Institut d'Estudis Espacials de Catalunya (IEEC), 08034 Barcelona, Spain\\
$^{12}$Institute of Cosmology and Gravitation, University of Portsmouth, Dennis Sciama Building, Portsmouth, PO1 3FX, UK\\
$^{13}$Institute of Space Sciences, ICE-CSIC, Campus UAB, Carrer de Can Magrans s/n, 08913 Bellaterra, Barcelona, Spain\\
$^{14}$Fermi National Accelerator Laboratory, PO Box 500, Batavia, IL 60510, USA\\
$^{15}$Center for Cosmology and AstroParticle Physics, The Ohio State University, 191 West Woodruff Avenue, Columbus, OH 43210, USA\\
$^{16}$Department of Physics, The Ohio State University, 191 West Woodruff Avenue, Columbus, OH 43210, USA\\
$^{17}$The Ohio State University, Columbus, 43210 OH, USA\\
$^{18}$School of Mathematics and Physics, University of Queensland, 4072, Australia\\
$^{19}$Sorbonne Universit\'{e}, CNRS/IN2P3, Laboratoire de Physique Nucl\'{e}aire et de Hautes Energies (LPNHE), FR-75005 Paris, France\\
$^{20}$NSF NOIRLab, 950 N. Cherry Ave., Tucson, AZ 85719, USA\\
$^{21}$Instituci\'{o} Catalana de Recerca i Estudis Avan\c{c}ats, Passeig de Llu\'{\i}s Companys, 23, 08010 Barcelona, Spain\\
$^{22}$Department of Physics and Astronomy, Siena College, 515 Loudon Road, Loudonville, NY 12211, USA\\
$^{23}$Department of Physics \& Astronomy and Pittsburgh Particle Physics, Astrophysics, and Cosmology Center (PITT PACC), University of Pittsburgh, 3941 O'Hara Street, Pittsburgh, PA 15260, USA\\
$^{24}$Departamento de F\'{i}sica, Universidad de Guanajuato - DCI, C.P. 37150, Leon, Guanajuato, M\'{e}xico\\
$^{25}$Instituto Avanzado de Cosmolog\'{\i}a A.~C., San Marcos 11 - Atenas 202. Magdalena Contreras, 10720. Ciudad de M\'{e}xico, M\'{e}xico\\
$^{26}$Instituto de Astrof\'{i}sica de Andaluc\'{i}a (CSIC), Glorieta de la Astronom\'{i}a, s/n, E-18008 Granada, Spain\\
$^{27}$Departament de F\'isica, EEBE, Universitat Polit\`ecnica de Catalunya, c/Eduard Maristany 10, 08930 Barcelona, Spain\\
$^{28}$Department of Astronomy, The Ohio State University, 4055 McPherson Laboratory, 140 W 18th Avenue, Columbus, OH 43210, USA\\
$^{29}$Department of Physics and Astronomy, Sejong University, Seoul, 143-747, Korea\\
$^{30}$CIEMAT, Avenida Complutense 40, E-28040 Madrid, Spain\\
$^{31}$Department of Physics, University of Michigan, Ann Arbor, MI 48109, USA\\
$^{32}$University of Michigan, Ann Arbor, MI 48109, USA\\
$^{33}$National Astronomical Observatories, Chinese Academy of Sciences, A20 Datun Rd., Chaoyang District, Beijing, 100012, P.R. China
}


\onecolumn
\appendix

\section{modelling Derivation}\label{appendix:modeling}
\input{derivations/model_derivation}


\bsp	
\label{lastpage}
\end{document}

%% file: derivations/model_derivation.tex
To compute the expectation value of the cross-correlation between projected shapes and the 3D matter field, $\mathcal{E}_{\rm model}$, we begin with their definitions, as described in Equations \ref{eq:E_complex} and \ref{eq:Q}:
\begin{equation}
    \epsilon = \tau [T_{xx} - T_{yy} + 2iT_{xy}]
\end{equation}
\begin{equation}
    Q(R_{\rm bin}, \pm\Pi_{\rm max}) = \frac{\int d^3r W(\bar{r})\delta e^{2i\theta_r}}{\int d^3r W(\bar{r})(1 + \xi_{\epsilon g})}
\end{equation}

\noindent Using these, $\mathcal{E}_{\rm model}$ is computed as:
\begin{equation}
    \begin{split}
    \mathcal{E}_{\rm model} = \Re \langle \epsilon*Q \rangle 
    = \frac{1}{2}\langle \epsilon^* Q + \epsilon Q^* \rangle = \Re\epsilon\Re Q + \Im\epsilon\Im Q = |Q|[(\epsilon_{xx} - \epsilon_{yy})\cos{2\theta} + 2\epsilon_{xy}\sin{2\theta}] \\
    = \frac{-\tau}{\int d^3r W(\bar{r})(1 + \xi_{\epsilon g})}
    \int dz \int RdR \int d\theta W(\bar{r})\delta(R, z)
        [(T_{xx} - T_{yy})\cos{2\theta} + (2T_{xy} - T_{xx} - T_{yy})\sin{2\theta}]
    \end{split}
\end{equation}

The 3 dimensions we integrate over here are the projected angle on the plane of the sky $\theta$, the projected distance, $R$, and the redshift $z$. $*$ indicates complex conjugation and $x^*$ is the complex conjugate of $x$. To compute the second integral, we convert to Fourier space.
\begin{equation}
    \begin{split}
        \int dz \int RdR W(\bar{r})\int d\theta \int \frac{d^3k}{(2\pi)^3} \tilde{\delta}(k) e^{-ik\cdot r}
        \int \frac{d^3q}{(2\pi)^3} e^{iq\cdot(0)}\tilde\delta{q}\frac{1}{q^2}
        [(q_xq_x - q_yq_y)\cos{2\theta} + (2q_xq_y - q_xq_x - q_yq_y + \frac{2}{3}q^2)\sin{2\theta}] \\
        = \int dz \int RdR W(\bar{r}) \int \frac{dk_z}{2\pi} \int \frac{KdK}{(2\pi)^2} P(k) \int d\phi \int d\theta e^{-iK\cdot R - ik_z z}
        \frac{1}{k^2} \\
        \big[K^2(\cos^2\phi - \sin^2\phi)\cos{2\theta} + K^2(2\cos\phi\sin\phi - (\cos^2\phi + sin^2\phi))\sin{2\theta} + \frac{2}{3}k^2\sin{2\theta}\big]\\
    \end{split}
\end{equation}

\noindent $\tilde{\delta}$ is the fractional overdensity in Fourier space. $k$ represented the 3D position in Fourier space, $K$ represents the 2D position on the plane of the sky ($k_x$, $k_y$), and $k_z$ along the line of sight. $k^2 = K^2 + k_z^2$. We then use the plane wave expansion $e^{iK\cdot R} = \sum_{n=-\infty}^{\infty}i^nJ_n(KR)e^{in\psi}$, where $\cos{\psi} = \hat{K}\cdot\hat{R}$. The above expression will integrate to $0$ for all $n$ except $n=\pm2$, allowing us to reduce $e^{iK\cdot R}$ to $-2J_2(KR) e^{2i(\theta-\phi)}$, of which the real component is $-2J_2(KR)\cos({2(\phi-\theta)})$.
The inner integrands becomes:
\begin{equation}
    \begin{split}
    -\int_0^{2\pi} d\phi \int_0^{2\pi} d\theta 2J_2(KR)\cos{2(\phi-\theta)} \big[K^2\cos{2\phi}\cos{2\theta} + K^2(\sin{2\phi}-1)\sin{2\theta} + \frac{2}{3}(K^2 + k_z^2)\sin{2\theta}\big]= -4\pi^2 J_2(KR)K^2
    \end{split}
\end{equation}
This leads to our final expression,

\begin{equation}
    \mathcal{E}_{\rm model} = \frac{-\tau}{\int d^3r W(\bar{r})(1 + \xi_{\epsilon g})}
    \int d^3r W(\bar{r})
    \int \frac{dk_z}{\text{2}\pi} \int KdK J_2(KR)\frac{K^2}{k^2}P_{gm}(k)e^{ik_z z}
\end{equation}

This is solved numerically, except for 
\begin{equation}
    \mathcal{J}_2(K) = \frac{2}{(R_{\rm max}^2 - R_{\rm min}^2)} \int_{R_{\rm min}}^{R_{\rm max}} RdR J_2(KR),
\end{equation}
for which we use the analytic solution:

\begin{equation}\label{eq:fancyJ_solution}
    \mathcal{J}_2(K, R) = \frac{2}{(R_\text{max}^2 - R_\text{min}^2)}\frac{1}{K^2} \big[ 2J_0(KR_\text{min}) + KR_\text{min}J_1(KR_\text{min}) - 2J_0(KR_\text{max}) - KR_\text{max}J_1(KR_\text{max}) \big]
\end{equation}